\documentclass[%
 reprint,
twocolumn, 
 amsmath,amssymb,
 aps,
prl,
superscriptaddress
]{revtex4-2}

\usepackage{graphicx}
\usepackage{dcolumn}
\usepackage{bm}
\usepackage[colorlinks=true, allcolors=blue,breaklinks]{hyperref}
\usepackage{braket}
\usepackage{bbding}
\usepackage{natbib}
\usepackage{gensymb}
\usepackage{balance}
\usepackage{pifont}
\usepackage{longtable}
\usepackage{times}
\usepackage{microtype}

\usepackage{amsmath,amssymb,amsfonts,mathrsfs}
\usepackage{CJK}
\usepackage{bm}
\usepackage{booktabs}
\usepackage{multirow}
\usepackage{longtable}
\usepackage{epstopdf}
\usepackage{dcolumn} 
\usepackage{babel}
\usepackage{graphicx}
\usepackage{float}
\allowdisplaybreaks
\usepackage{braket}
\usepackage{color}
\usepackage{makecell}
\usepackage{bbding}
\usepackage{utfsym}
\usepackage{pifont}
\usepackage{graphicx}
\usepackage{dcolumn}
\usepackage{bm}
\usepackage[colorlinks=true, allcolors=blue]{hyperref} 

\begin{document}


\title{Odd-Parity Chiral Magnons in Collinear Antiferromagnetic Multiferroics}

\author{Quanchao Du}
\email{These authors contributed equally to this work.}
\affiliation{%
 Ministry of Education Key Laboratory for Nonequilibrium Synthesis and Modulation of Condensed Matter, Shaanxi Province Key Laboratory of Advanced Functional Materials and Mesoscopic Physics, School of Physics, Xi’an Jiaotong University, Xi’an 710049, China 
}%

\author{Zhenlong Zhang}
\email{These authors contributed equally to this work.}
\affiliation{%
 Ministry of Education Key Laboratory for Nonequilibrium Synthesis and Modulation of Condensed Matter, Shaanxi Province Key Laboratory of Advanced Functional Materials and Mesoscopic Physics, School of Physics, Xi’an Jiaotong University, Xi’an 710049, China 
}%

\author{Yuanjun Jin}
\affiliation{%
 Guangdong Basic Research Center of Excellence for Structure and Fundamental Interactions of Matter,
 Guangdong Provincial Key Laboratory of Quantum Engineering and Quantum Materials,
 School of Physics, South China Normal University, Guangzhou 510006, China 
} 

\author{Rui Li}
\affiliation{%
 Ministry of Education Key Laboratory for Nonequilibrium Synthesis and Modulation of Condensed Matter, Shaanxi Province Key Laboratory of Advanced Functional Materials and Mesoscopic Physics, School of Physics, Xi’an Jiaotong University, Xi’an 710049, China 
}%

\author{Haibo Xie}
\affiliation{%
 Ministry of Education Key Laboratory for Nonequilibrium Synthesis and Modulation of Condensed Matter, Shaanxi Province Key Laboratory of Advanced Functional Materials and Mesoscopic Physics, School of Physics, Xi’an Jiaotong University, Xi’an 710049, China 
}%

\author{Zhe Wang}
\affiliation{%
 Ministry of Education Key Laboratory for Nonequilibrium Synthesis and Modulation of Condensed Matter, Shaanxi Province Key Laboratory of Advanced Functional Materials and Mesoscopic Physics, School of Physics, Xi’an Jiaotong University, Xi’an 710049, China 
}%

\author{Zhijun Jiang}
\email{zjjiang@xjtu.edu.cn}
\affiliation{%
 Ministry of Education Key Laboratory for Nonequilibrium Synthesis and Modulation of Condensed Matter, Shaanxi Province Key Laboratory of Advanced Functional Materials and Mesoscopic Physics, School of Physics, Xi’an Jiaotong University, Xi’an 710049, China 
}%

\author{Lei Zhang}
\affiliation{%
 Ministry of Education Key Laboratory for Nonequilibrium Synthesis and Modulation of Condensed Matter, Shaanxi Province Key Laboratory of Advanced Functional Materials and Mesoscopic Physics, School of Physics, Xi’an Jiaotong University, Xi’an 710049, China 
}

\author{Hongjian Zhao}
\affiliation{%
 Key Laboratory of Material Simulation Methods and Software of Ministry of Education,
College of Physics, Jilin University, Changchun 130012, China
}%

\author{Jinyang Ni}
\email{jyni@xjtu.edu.cn}
\affiliation{%
 Ministry of Education Key Laboratory for Nonequilibrium Synthesis and Modulation of Condensed Matter, Shaanxi Province Key Laboratory of Advanced Functional Materials and Mesoscopic Physics, School of Physics, Xi’an Jiaotong University, Xi’an 710049, China 
}%




\begin{abstract}
Odd-parity magnetism represents an intriguing frontier in unconventional magnets; however, its realization has traditionally relied on noncollinear magnetic orders accompanied by broken spin conservation, which inevitably causes spin relaxation and dissipative transport of spin-encoded information. Here, we uncover a distinct route toward spin-conserving odd-parity magnon splitting enabled by antichiral Haldane-like flux (AHF) in collinear antiferromagnetic multiferroics. Symmetry analysis reveals that such AHF driven by intra-sublattice Dzyaloshinskii-Moriya interactions (DMI) gives rise to symmetry-dependent odd-parity magnon splittings, ranging from $p$-wave and $f$-wave to nodeless forms. Moreover, the coupling between ferroelectric modes and DMI provides an electric-field knob for manipulating chiral band splitting and magnon transport. Combining density functional theory (DFT) calculations, we identify a series of promising candidate materials in both 2D and bulk antiferromagnetic multiferroics. Our work provides new insights into the realization of odd-parity chiral magnons in collinear antiferromagnets and magnetoelectric coupling mechanisms in multiferroics.
\end{abstract}

\maketitle
\newpage

\textit{Introduction.} Magnons are the quanta of collective spin excitations in ordered magnets\,\cite{chumak2014magnon, chumak2015magnon}. Owing to their charge neutrality, magnons can propagate through insulators without Joule heating, offering a promising route towards lossless spintronics\,\cite{ chumak2014magnon, chumak2015magnon, onose2010observation, bader2010spintronics, mcclarty2022topological,lenk2011building,  chen2025unconventional, chen_PRX_2018_8, mook_PRX_2021_11, wang2019nonreciprocity, bai2026ferroelectrics}. Compared with ferromagnets (FMs), magnons in antiferromagnets (AFMs) exhibit much higher intrinsic frequencies, reaching the terahertz regime, while naturally hosting both \textit{left}- and \textit{right}-handed chirality that can be exploited for spin-information encoding, storage and processing\,\cite{baltz2018antiferromagnetic, jungwirth2016antiferromagnetic, rezende2019introduction}. Generally, in conventional AFMs, opposite-spin sublattices are related by a combination of translation and inversion symmetries, enforcing magnon band degeneracy. Recent studies have shown that when opposite-spin sublattices are instead connected by additional rotational or mirror symmetries, collinear magnets can host spin-conserving magnon chiral splittings with even-parity momentum dependence\,\cite{vsmejkal2020crystal, mazin2021prediction, bai2024altermagnetism, zhou2025manipulation,  duan2025antiferroelectric, song2025altermagnets, liu2022spin, vsmejkal2023chiral, hoyer2025altermagnetic, cui2023efficient, liu2024chiral, sun2025observation, singh2025chiral, morano2025absence}. However, even-parity spin splittings remain invariant under the effective inversion symmetry, limiting their electrical controllability\,\cite{song2025electrical,priessnitz2026ferroelectric, neumann2026odd, LuoOdd, 7kmk-yl2t, c8pd-2fs4}. In contrast, odd-parity spin splitting can reverse sign under inversion symmetry, offering a natural route to electrical control; yet, their realization has so far been limited in noncollinear magnetic orders\,\cite{song2025electrical, priessnitz2026ferroelectric, neumann2026odd, LuoOdd, 7kmk-yl2t, c8pd-2fs4}. Noncollinear magnets generally exhibit low N\'{e}el temperature, while the accompanying breaking of spin conservation can introduce additional spin dissipation, imposing fundamental constraints on the stability and coherent transport of magnons\,\cite{pirro2021advances, hortensius2021coherent, chen2025observation}.

In this work, we uncover a symmetry-governed route toward spin-conserving odd-parity chiral magnons in collinear antiferromagnetic multiferroics. We demonstrate that intra-sublattice Dzyaloshinskii-Moriya interaction (DMI) induces an effective antichiral Haldane-like flux (AHF), producing $f$-wave chiral magnon splitting in hexagonal multiferroics. We further establish symmetry principles governing their $p$-wave and nodeless forms. Importantly, such odd-parity magnons enable ferroelectrically switchable spin transport, offering a new pathway for magnetoelectric (ME) coupling of charge-neutral quasi-particles. Combining symmetry analysis with DFT calculations, we identify the possible magnetic space groups and a broad family of 2D and bulk multiferroic candidates.

\begin{figure}
    	\centering
    	\includegraphics[scale=0.5]{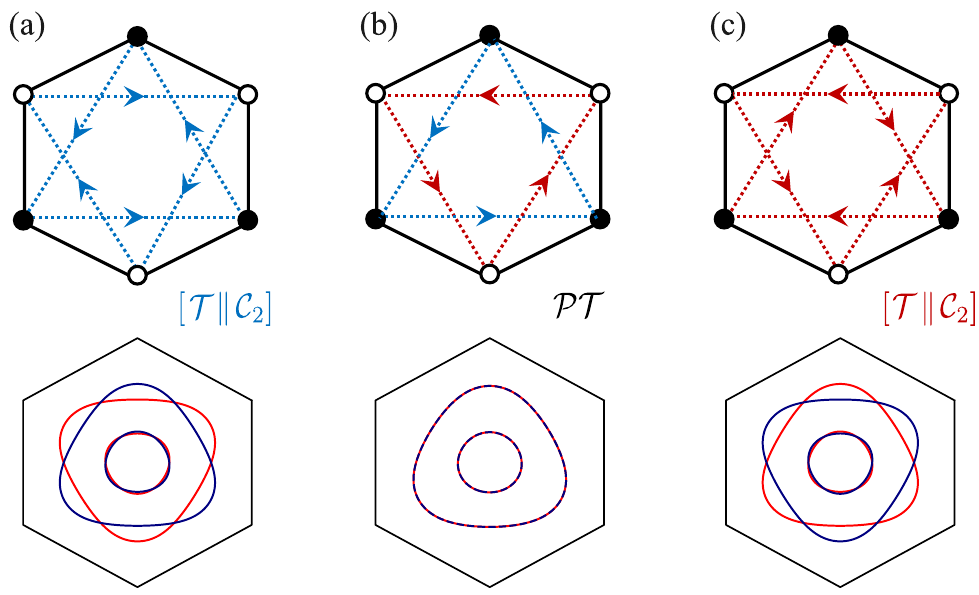}
    \caption{(a) AHF phase in collinear AFMs with opposite-spin sublattices connected by the $[{\cal T}\parallel{\cal C}_{2}]$.
(b) HF phase in ${\cal PT}$-symmetric AFMs.  
(c) The reversed configuration of (a). Here, intra-sublattice DMI vectors along the $-z$ and $+z$
directions are indicated by red and blue lines, respectively. The lower panels show the band splitting between the $\alpha$ (red) and $\beta$ (blue) magnon modes.}
    \label{fig1}
\end{figure}

\textit{Symmetry analysis}.
We first establish the symmetry constraints governing the chirality of intra-sublattice DMI vectors in honeycomb N\'{e}el AFMs. As shown in Fig.\,\ref{fig1}(b), in easy-axis AFMs with opposite-spins related by the combined ${\cal PT}$ symmetry, where ${\cal P}$ (${\cal T}$) denotes inversion (time-reversal) symmetry, intra-sublattice DMI vectors on opposite-spin sublattices acquire opposite signs, forming the Haldane-like flux (HF) phase\,\cite{cheng2016spin, zyuzin2016magnon, 9yz2-mqd9,du2025nonreciprocal, ni2025magnon}. By contrast, when the opposite-spin sublattices are related by a two-fold screw axis ${\cal C}_2$ along the $z$ axis combined with ${\cal T}$, denoted as $\left[{\cal T}\parallel {\cal C}_2\right]$, or by a vertical mirror plane ${\cal M}$ combined with a two-fold spin rotation ${\cal C}_{2\perp}$ about an axis perpendicular to the N\'eel vector, denoted as $\left[{\cal C}_{2\perp}\parallel {\cal M}\right]$, the intra-sublattice DMI vectors likewise acquire identical signs\,\cite{rn1l-d6cq, liu20252D, 9yz2-mqd9}. In this case, as illustrated in Fig.\,\ref{fig1}(a), the intra-sublattice DMI vectors generate an AHF phase, which is fundamentally distinct from the HF phase protected by ${\cal PT}$ symmetry. Notably, both the HF and AHF phases preserve sublattice symmetry, which ensures band degeneracy in the nonrelativistic limit.

\begin{figure*}
\includegraphics[width=0.75\textwidth]{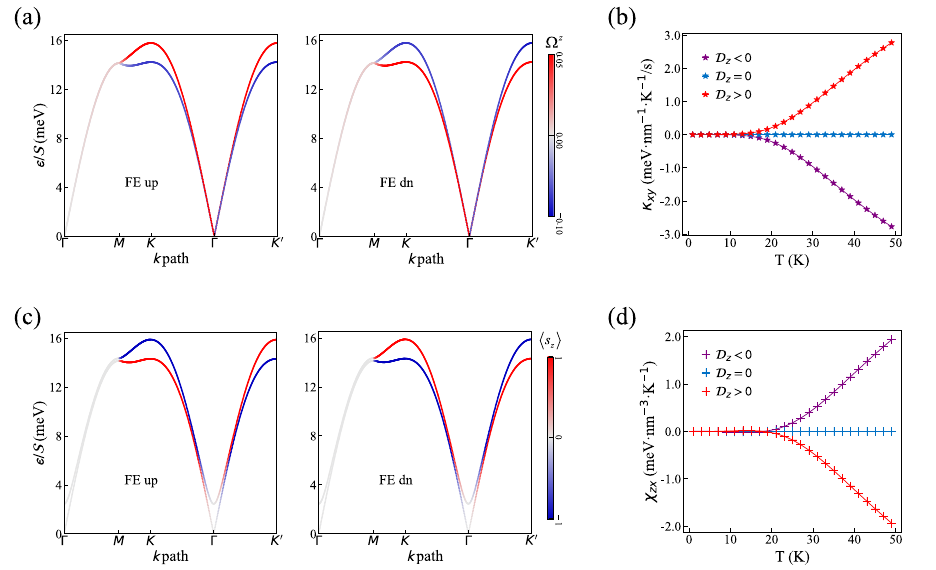}
\caption{\label{fig2} (a) $f$-wave chiral magnon bands weighted by $\Omega^{z}$ for different ferroelectric states (${\cal D}_{z}$\,$>$\,0 and ${\cal D}_{z}$\,$<$\,0). (b) The calculated magnon Hall conductivity $\kappa_{xy}$ with different ${\cal D}_{z}$, where ${\cal J}_{1}$\,$=$\,$\mbox{5}$\,$\mbox{meV}$ and ${\cal D}_{z}$\,$=$\,$\pm$\,$\mbox{0.15}$\,$\mbox{meV}$. (c) The magnon bands weighted by the $\langle{s_z}\rangle$ in easy-plane antiferromagnetic lattice for different ferroelectric states. (d) The calculated Edelstein response parameter $\chi_{zx}$ with different ${\cal D}_{z}$.}
\end{figure*}

\textit{Model for $f$- and $p$-wave magnons}.
Building on the above symmetry analysis, the minimal spin model for such collinear honeycomb AFMs can be expressed as\,:
\begin{equation} \label{eq1}
{\cal \hat{H}} = {\cal J}_{1} \sum_{\langle {i,j}\rangle}  {{\cal S}_{i} \cdot{\cal S}_{j}} + {\cal D}_{z}\sum_{\langle i,j\rangle^{\prime}} {{\boldsymbol{\nu_{ij}}}\cdot\left({\cal S}_{i}\times{\cal S}_{j} \right)}, 
\end{equation}
where the first term describes the NN Heisenberg inter-sublattice antiferromagnetic exchange, and the second term is the $z$ component DMI. For ${\cal PT}$-symmetric case, ${\nu}_{ij}$\,$=$\,$\pm1$, such that the DMI changes sign when it connects two spin-up sites or two spin-down sites\,[see Fig.\,\ref{fig1}(b)]\,\cite{cheng2016spin, zyuzin2016magnon, 9yz2-mqd9,du2025nonreciprocal, ni2025magnon}. While opposite spin sublattices are connected by the $\left[{\cal C}_{2\perp}\parallel {\cal M}\right]$ or $\left[{\cal T}\parallel {\cal C}_2\right]$ symmetry, DMI acts equally in both spin-up and spin-down sublattices: ${\nu}_{ij}$\,$=$\,$+1$ (or ${\nu}_{ij}$\,$=$\,$-1$) for all sites. In the following, we compare the magnon bands of these two cases.  

After Holstein-Primakoff and linear spin wave approximation (LSW)\,\cite{holstein1940field, ni2025magnon}, the quadratic magnon Hamiltonian of Eq.\,(\ref{eq1}) can be expressed in the basis $\psi^{\dagger}_{\textbf{k}} = (\hat{a}_{\textbf{k}}^{\dagger}, \hat{b}_{-\textbf{k}}, \hat{a}_{-\textbf{k}}, \hat{b}^{\dagger}_{\textbf{k}})$ as ${\cal \hat{H}} = \sum_{\textbf{k}}\psi^{\dagger}_{\textbf{k}} {\cal \hat{H}}_{\textbf{k}} \psi_{\textbf{k}}$. For the ${\cal PT}$-symmetric AFMs, ${\cal \hat{H}}_{\textbf{k}}$ can be expressed as\,:
\begin{equation} {\label{eq2}}
\frac{{\cal \hat{H}}_{\textbf{k}}}{S} = \lambda I + \left(\begin{matrix}
   {f_{\textbf{k}}} &\gamma_{\textbf{k}} & 0 & 0\\
    \gamma_{\textbf{k}}^{\dagger}& -{f_{\textbf{k}}} &0 & 0 \\
    0 & 0 & -{f_{\textbf{k}}} &  \gamma_{\textbf{k}}^{\dagger} \\
    0 & 0 & \gamma_{\textbf{k}} & {f_{\textbf{k}}}
\end{matrix} \right),
\end{equation}
where $\gamma_{\textbf{k}}$\,$=$\,${\cal J}_{1}\sum_{i} \mbox{exp}(i \mathbf{k}\cdot \boldsymbol{\delta}_{i})$, $f_{\textbf{k}}$\,$=$\,${\cal D}_{z}\sum_{i\in odd}2\mbox{sin}(\mathbf{k}\cdot \boldsymbol{\mu}_{i})$, $\lambda$\,$=$\,$3{\cal J}_{1}$. Here, $\boldsymbol{\delta}_{i}(\boldsymbol{\mu}_{i})$ denotes the inter-sublattice (intra-sublattice) linking vectors, as illustrated in Figs.\,S1(a) and (b) of Supplemental Materials\,(SM)\,\cite{Supplemental_Materials}. By employing the Bogoliubov transformation\,\cite{bogoljubov1958new,valatin1958comments}, the eigenvalues of Eq.\,(\ref{eq2}) can be obtained\,: 
\begin{equation} {\label{eq3}}
\begin{split}
        \epsilon_{\alpha,\beta}(\textbf{k}) & = \sqrt{\lambda^{2} - |\gamma_{\textbf{k}}|^{2}} + f_{\textbf{k}}. 
\end{split}
\end{equation}
Clearly, as shown in Fig.\,\ref{fig1}(b), the $\alpha$ and $\beta$ magnon modes are degenerate throughout the Brillouin zone, and the emergence of ${\cal D}_{z}$ only induces nonreciprocity between $\textbf{k}$ and $-\textbf{k}$ points\,\cite{du2025nonreciprocal, ni2025magnon, matsumoto2020nonreciprocal, sato2019nonreciprocal, gitgeatpong2017nonreciprocal}. 

When DMI acts equally in both spin-up and spin-down sublattices, corresponding to the AHF phase, ${\cal \hat{H}}_{\textbf{k}}$ is modified as\,:
\begin{equation} {\label{eq4}}
 \frac{{\cal \hat{H}}_{\textbf{k}}}{S} = \lambda  I + \left(\begin{matrix}
   {f_{\textbf{k}}} &\gamma_{\textbf{k}} & 0 & 0\\
    \gamma_{\textbf{k}}^{\dagger}& {f_{\textbf{k}}} &0 & 0 \\
    0 & 0 & -{f_{\textbf{k}}} &  \gamma_{\textbf{k}}^{\dagger} \\
    0 & 0 & \gamma_{\textbf{k}} & -{f_{\textbf{k}}}
\end{matrix} \right).
\end{equation}
The corresponding eigenvalues are given as\,:
\begin{equation}\label{eq5}
        \epsilon_{\alpha, \beta}(\textbf{k})  = \sqrt{(\lambda  \pm f_{\textbf{k}})^{2} - |\gamma_{\textbf{k}}|^{2}}. 
\end{equation}
The introduction of ${\cal D}_{z}$ lifts the degeneracy between $\alpha$ and $\beta$ modes, except along the ${\Gamma}$ to $\mbox{M}$ path, giving rise to the $f$-wave magnon band splitting, as shown in Fig.\,\ref{fig1}(a) and Fig.\,\ref{fig2}(a). The magnon band splitting, $\Delta(\textbf{k})$\,$=$\,$\epsilon_{\alpha}(\textbf{k})$\,$-$\,$\epsilon_{\beta}(\textbf{k})$, reaches its maximum at the $\mbox{K}$ point, with a magnitude of approximately $6\sqrt{3}\,{\cal D}_{z}$\,\cite{Supplemental_Materials}. To elucidate the nature of the $f$-wave band splitting, we expand the $f_{\textbf{k}}$ near the ${\Gamma}$ point as $f_{\textbf{k}}$\,$\propto$\,$k_{y}(3k^{2}_{x}-k^{2}_{y})$. Accordingly, the low-energy band splitting scales as $\Delta(\textbf{k})$\,$\approx$\,$2f_{\textbf{k}}$\,$\propto$\,$k^{3}$, manifesting the symmetry-protected $f$-wave odd-parity characteristics. Note that this phase also allows a finite inter-sublattice DMI, but it does not contribute to the magnon band splitting due to the ${\cal C}_{3}$ symmetry of honeycomb lattice\,\cite{Supplemental_Materials}.

\begin{figure}
    	\includegraphics[scale=0.5]{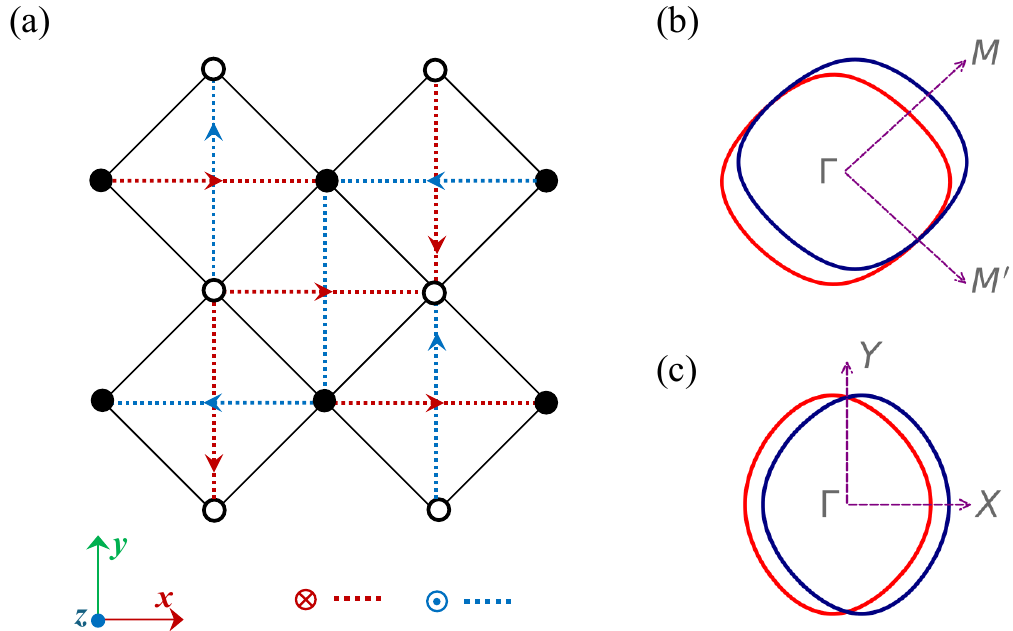}
    	\caption{(a) Top view of orthorhombic AFMs with intra-sublattice DMI forming the AHF. White and black dots denote spin-up and spin-down sites, respectively. (b) $p$-wave chiral splitting in square-lattice ($\mathcal{C}_{4}$) AFMs. (c) $p$-wave chiral splitting in rectangular-lattice  ($\mathcal{C}_{2}$) AFMs. }
    \label{fig3}
\end{figure}

The above results can be naturally extended to orthorhombic AFMs with N\'{e}el order. As shown in Fig.\,\ref{fig3}(a) and Fig.\,S5\,\cite{Supplemental_Materials}, when opposite-spin sublattices are related by the $\left[{\cal T}\parallel {\cal C}_2\right]$ or $\left[{\cal C}_{2\perp}\parallel {\cal M}\right]$ symmetry, the inter-sublattice DMI forms an AHF phase, thereby enabling a $p$-wave magnon band splitting. The crossing nodes associated with the $p$-wave splitting are further constrained by lattice symmetry. For AFMs with ${\cal C}_{4}$ rotational symmetry, namely, the square lattice, as illustrated in Fig.\,\ref{fig3}(b), the maximal splitting occurs along the $\Gamma$--$\mathrm{M}$ direction, whereas the twofold degeneracy is preserved along the $\Gamma$--$\mathrm{M}^{\prime}$ direction. Upon further reducing the rotational symmetry to ${\cal C}_{2}$, corresponding to the rectangular lattice, the maximal splitting shifts to the $\Gamma$--$\mathrm{X}$ direction, while the bands remain degenerate along the $\Gamma$--$\mathrm{Y}$ path [see Fig.\,\ref{fig3}(c)]. To further distinguish between the two cases, we expand $f_{\mathbf{k}}$ at low energies for the square and rectangular lattices, obtaining $f_{\mathbf{k}}$\,$\propto$\,$k_x$\,$-$\,$k_y$ and $f_{\mathbf{k}}$\,$\propto$\,$k_x$, respectively, both of which exhibit the characteristic linear momentum dependence of $p$-wave splitting while showing a pronounced path dependence.

Unlike the $f$-wave chiral magnons, the $p$-wave splitting shifts the band minimum away from the $\Gamma$ point to two symmetric points in its vicinity, resulting in a Rashba-like dispersion \cite{kawano2019designing,okuma2017magnon}. Additionally, for orthorhombic AFMs, the symmetry-allowed inter-sublattice DMI can also induce a sizable $p$-wave magnon splitting\,\cite{Supplemental_Materials}. Despite these differences, both $f$-wave and $p$-wave magnon Hamiltonians remain block diagonal, thereby preserving spin conservation. Consequently, the split $\alpha$ and $\beta$ modes carry opposite $\left<s_z\right>$, with chirality of $\mp 1$, respectively\,\cite{cheng2016spin, zyuzin2016magnon, 9yz2-mqd9,du2025nonreciprocal, ni2025magnon}.  Another key feature of such odd-parity chiral magnons is that the chiral splitting can be reversibly switched by reversing the DMI vector\,\cite{dzyaloshinsky_JPCS_1958_4,moriya_PR_1960_120, zhao2021dzyaloshinskii, zhao2014near}, as shown in Fig.\,\ref{fig2}(a). This indicates that, for collinear antiferromagnetic multiferroics satisfying the above symmetry constraints, the ferroelectric polarization is directly coupled to the intra-sublattice DMI, enabling ferroelectric control of odd-parity chiral magnons.

\textit{Magnon transport}.
Beyond the magnon bands, HF and AHF phases are also manifested in their corresponding magnon eigenvectors. In AFMs hosting the HF phase, the intra-sublattice ${\cal D}_z$ modifies the magnon dispersion solely by introducing nonreciprocity, while leaving the eigenvectors essentially unchanged. In contrast, in AHF phase, ${\cal D}_z$ substantially modifies the distribution of the magnon eigenvectors, as detailed in Sec.\,II of SM\,\cite{Supplemental_Materials}. Therefore, the HF and AHF phases in AFMs give rise to qualitatively distinct magnon quantum geometry, leading to different Berry curvature driven magnon transport behaviors\,\cite{xiao2010berry}.

As detailed in the Sec.\,II of SM\,\cite{Supplemental_Materials}, we derive the $z$ component magnon Berry curvature $\Omega^{z}$ in the HF and AHF phases of N\'eel AFMs with honeycomb lattice. The above analysis shows that, in the HF phase, ${\cal D}_z$ is decoupled from the magnon eigenvectors, and $\Omega^{z}$ is therefore independent ${\cal D}_z$, taking the following form
\begin{equation}\label{eq6}
\Omega_{\alpha, \beta}^{z}(\textbf{k})  = \mp \frac{\lambda}{2 {{\epsilon_{0}}^3}}\left(\nabla \operatorname{Re} \gamma_{\textbf{k}} \times \nabla \operatorname{Im} \gamma_{\textbf{k}}\right),
\end{equation}
with $\epsilon_{0}$\,$=$\,$\sqrt{\lambda^{2} - |\gamma_{\textbf{k}}|^{2}}$. In this case, the distribution of $\Omega^{z}$ satisfies the valley symmetry, $\mbox{i.e.}$, $\Omega^{z}(\textbf{k})$\,$=$\,$-\Omega^{z}(-{\textbf{k}})$\,\cite{brinkman1966theory, ni2025magnon}, as depicted in Fig.\,S3\,\cite{Supplemental_Materials}. Given the band degeneracy, the anomalous Hall effect (AHE) of magnons vanishes, and only the spin Nernst effect (SNE) still exists upon applying the temperature gradient\,\cite{cheng2016spin, zyuzin2016magnon, ni2025magnon}. In contrast, the magnon Berry curvature of AHF phase becomes dependent on ${\cal D}_{z}$ and can be expressed as\,:
\begin{equation}\label{eq6}
\begin{aligned}
& \Omega_{\alpha, \beta}^{z}(\textbf{k})  = \mp \frac{1}{2 {{\epsilon^{\prime}_{0}}^3}}\Big[\left(\lambda+f_{\textbf{k}}\right)\left(\nabla \operatorname{Re} \gamma_{\textbf{k}} \times \nabla \operatorname{Im} \gamma_{\textbf{k}}\right) \\
& +\operatorname{Im} \gamma_{\textbf{k}}\left(\nabla f_{\textbf{k}} \times \nabla \operatorname{Re} \gamma_{\textbf{k}}\right) +\operatorname{Re} \gamma_{\textbf{k}}\left(\nabla \operatorname{Im} \gamma_{\textbf{k}} \times \nabla f_{\textbf{k}}\right)\Big],
\end{aligned}
\end{equation}
where $\epsilon^{\prime}_{0}$\,$=$\,$\sqrt{\left(\lambda + f_{\textbf{k}}\right)^{2} - |\gamma_{\textbf{k}}|^{2}}$. The emergence of ${\cal D}{z}$ breaks the valley symmetry of the Berry curvature, $\Omega^{z}(\mathbf{k})\neq-\Omega^{z}(-\mathbf{k})$, and significantly reshapes the Berry curvature near the $\mbox{K}$ and $\mbox{K}^{\prime}$ valleys, as shown in Fig.\,\ref{fig2}(a) and Fig.\,S3\,\cite{Supplemental_Materials}. The giant chiral magnon splitting and asymmetric Berry curvature collectively contribute to anomalous Hall transport\,\cite{matsumoto2011theoretical, onose2010observation}. Notably, reversing DMI vector can reverse the Berry curvature, which renders the magnon Hall response ferroelectrically switchable [see Fig.\,\ref{fig2}(b)]. Importantly, as charge neutral quasi-particles, magnons do not require a Fermi surface for transport, avoiding the polarization instability caused by carrier accumulation in conventional electronic systems.

In addition to Berry-curvature-driven magnon transport, magnon band splitting can give rise to longitudinal spin currents, such as the magnon spin Seebeck effect (SSE)\,\cite{cui2023efficient, wu2025magnon, wu2025magnon}. However, the SSE is intrinsically forbidden in $f$-wave and $p$-wave chiral magnons due to their odd-parity band splitting. For spin accumulation induced by the Edelstein effect (EdE)\,\cite{libo2020,neumann2026odd} spin conservation must be broken; therefore, easy-plane magnetic anisotropy is generally required to relax this constraint.

\textit{Nodeless chiral magnons}.
Upon introducing easy-plane anisotropy, the Hamiltonian of Eq.\,(\ref{eq4}) is modified as 
\begin{equation} {\label{eq7}}
 \frac{{\cal \hat{H}}_{\textbf{k}}}{S} = \lambda  I + \left(\begin{matrix}
   {f_{\textbf{k}}} &\gamma_{\textbf{k}} & {\mathcal K}_{x} & 0\\
    \gamma_{\textbf{k}}^{\dagger}& {f_{\textbf{k}}} &0 & {\mathcal K}_{x} \\
    {\mathcal K}_{x} & 0 & -{f_{\textbf{k}}} &  \gamma_{\textbf{k}}^{\dagger} \\
    0 & {\mathcal K}_{x} & \gamma_{\textbf{k}} & -{f_{\textbf{k}}}
\end{matrix} \right),
\end{equation}
where $\lambda$ is modified as ${3\mathcal J}_{1}$\,$+$\,${\mathcal K}_{x}$, with ${\mathcal K}_{x}$\,$>$\,0 denoting the easy-plane anisotropy.
Compared to Eq.\,(\ref{eq4}), the emergence of ${\mathcal K}_{x}$ leads to the Hamiltonian not being block diagonal, thereby breaking spin conservation. By employing Bogoliubov transformation, the magnon bands are obtained as shown in Fig.\,\ref{fig2}(c), which demonstrate that ${\mathcal K}_{x}$ additionally lifts the band degeneracy along the $\Gamma$ to $\mbox{M}$ path. Unlike the aforementioned $f$-wave and $p$-wave cases, the corresponding magnon bands exhibit a nodeless character. Meanwhile, the spin-momentum texture of such magnon bands preserves its odd-parity character and exhibits ferroelectric switchability. 

The odd-parity spin-momentum locking ensures the spin accumulation driven by the magnon EdE including extrinsic contributions\,\cite{libo2020, neumann2026odd}. At equilibrium, magnons at $\textbf{k}$ and $-\textbf{k}$ possess opposite spin polarizations, yielding zero net spin accumulation. A temperature gradient breaks this balance by driving asymmetric magnon redistribution between $\textbf{k}$ and $-\textbf{k}$ states, resulting in a finite net spin polarization. The calculated response coefficient $\chi_{zx}$ as a function of temperature is shown in Fig.\,\ref{fig2}(d), demonstrating that DMI induces a sizable spin polarization with a ferroelectrically switchable character. Similar ferroelectric-switchable spin polarization effects also exist in orthorhombic antiferromagnetic multiferroics, as detailed in the Sec.\,III of SM\,\cite{Supplemental_Materials}.

\textit{Material candidates}. As mentioned, collinear N\'{e}el AFMs with opposite-spins sublattices related by $\left[{\cal T}\parallel {\cal C}_{2}\right]$ or $\left[{\cal C}_{2\perp}\parallel {\cal M}\right]$ symmetries host an AHF driven by inter-sublattice DMI, giving rise to $f$-wave and $p$-wave chiral magnon splitting. Through magnetic symmetry analysis, we identify the $6mm$, $6$, $4mm$, $4$ and $mm2$ magnetic point groups that satisfy the these symmetry conditions. In particular, the $6$ and $6mm$ magnetic point groups permit $f$-wave chiral magnons, for which wurtzite-type $\mbox{CoS(Se)}$ and hexagonal $\mbox{LuMnO}_{3}$ are promising multiferroic candidates\,\cite{grzybowski2024wurtzite,bezzerga2025high, van2004origin}. As depicted in Fig.\,S1(a)\,\cite{Supplemental_Materials}, $\mbox{CoS(Se)}$ and $\mbox{LuMnO}_{3}$ crystallize in the space group $P6_{3}mc$, where magnetic ions are ferromagnetically aligned within the $xy$ plane and antiferromagnetically coupled along the $z$ axis. DFT calculations reveal that $\mbox{CoS(Se)}$ and $\mbox{LuMnO}_{3}$ both exhibit easy-axis magnetic anisotropy with magnetic group $P6_3^\prime c m^\prime$ (see Tab.\,S1\,\cite{Supplemental_Materials}). For this easy-axis N\'eel order, the opposite-spin sublattices are related by $\left[{\cal T}\parallel {\cal C}_{2}\tau\right]$ or $\left[{\cal C}_{2\perp}\parallel {\cal M}\tau\right]$ with $\tau$ being half lattice translation along $z$ axis, which enforce the equal intra-sublattice ${\cal D}_{z}$ and allow their reversal through ferroelectric switching. With negligible inter-sublattice DMI, the $f$-wave magnon splitting is confined to the $k_x$–$k_y$ plane and is ferroelectrically switchable, as is the associated magnon transport. Notably, the $f$-wave magnon splitting in bulk $\mbox{CoSe}$ can reach approximately $10\,\mathrm{meV}$, making it readily accessible to neutron-scattering experiments\,\cite{liu2024chiral, sun2025observation, singh2025chiral, morano2025absence}.

For $p$-wave chiral magnons, we identify a class of orthorhombic antiferromagnetic multiferroics encompassing both bulk and 2D candidates, as summarized in Tab.\,\ref{tab1} and Fig.\,S1(d)\,\cite{Supplemental_Materials}. Bulk $\mbox{K}_{3}\mbox{Cr}_{2}\mbox{F}_{7}$\,\cite{xpsy-cjn6, xu2017designing} and monolayer $\mbox{VSI}_{2}$\,\cite{zhu2025two, Tan2019} possess easy-axis N\'eel order accompanied by pronounced chiral magnon splitting, with magnon gap exceeding 4\,$\mbox{meV}$. We further identify several candidates for nodeless chiral magnon, including $\mathrm{CoO}$, $\mathrm{MnS}$, and $\mathrm{MnSe}$ with easy-plane N\'eel order, as well as the orthorhombic antiferromagnet $\mathrm{VOBr}_{2}$. These candidates are all insulating, which ensures the stability of the polarization during ferroelectric switching.

\setlength{\tabcolsep}{1pt}
\renewcommand\arraystretch{2}
\renewcommand{\thetable}{\arabic{table}}

\begin{table*}[t]
\centering
\caption{The symmetry-allowed magnetic space groups for odd-parity chiral magnons. Here, we focus on layered collinear AFMs with the rotation axis along the $z$ direction. For $6mm$, $6$, $4mm$, and $4$, the N\'eel vectors ($\hat{\cal N}$) are along the $z$ axis. For $mm2$, $\hat{\cal N}$ can be either in-plane ($x/y$) or out-of-plane ($z$), as indicated in parentheses. The symmetry allowed magnon transport and material candidates are also listed.}
\label{tab1}
\scriptsize
\begin{tabular}{c|c|c|c|c}
\hline\hline
Point groups
& $6mm$, $6$\,\,($f$-wave)
& $4mm$, $4$\,\,($p$-wave)
& $mm2$\,\,($p$-wave)
& $mm2$\,\,(Nodeless) \\
\hline
\makecell[c]{Magnetic groups}
& \makecell[c]{$P6cc$, $P6_3^\prime m^\prime c$, $P6_3^\prime$, $P6_3^\prime c m^\prime$}
& \makecell[c]{$P4_2^\prime$, $P4_2^\prime nm^\prime$,
$P4_2^\prime cm^\prime$, \\ $P4_2^\prime c^\prime m$, $P4cc$, $P4nc$,\\
 $P4_2^\prime m^\prime c$, $P4_2 bc$,
$P4_2^\prime b^\prime c$, $P4_2^\prime$}
& \makecell[c]{$Pcc2$, $Pma2$, $Pnc2$, $Pba2$ \\ $Pnn2$, $Ccc2$, $Abm2$, $Aba2$}
& \makecell[c]{$Pmm2$, $Pmc2_1$, $Pma2$, $Pnc2$,\\ $Pmn2_1$, $Cmc2_1$, $Ama2$} \\
\hline

\makecell[c]{Magnon transport}
& \makecell[c]{AHE \ding{51}, SNE \ding{51}, SSE \ding{55}, EdE \ding{55}}
& \multicolumn{2}{c|}{\makecell[c]{AHE \ding{51}, SNE \ding{51}, SSE \ding{55}, EdE \ding{55}}}
& \makecell[c]{AHE \ding{55}, SNE \ding{55}, SSE \ding{55}, EdE \ding{51}}\\
\hline
\makecell[c]{Material candidates $\&$ \\ band splitting\,(meV) }
& \makecell[c]{$\mbox{CoS}$\,(6.5), $\mbox{CoSe}$\,(10), $\mbox{LuMnO}_{3}$\,(1.2)}
& \multicolumn{2}{c|}{\makecell[c]{$\mbox{K}_{3}\mbox{Cr}_{2}\mbox{F}_{7}$\,(4.2), $\mbox{Sr}_{2}\mbox{Mn}\mbox{O}_{2}\mbox{F}$\,(0.9), $\mbox{VSI}_{2}$\,(8.1), $\mbox{VSBr}_{2}$\,(3.7)}}
& \makecell[c]{$\mbox{MnSe}$\,(1.5), $\mbox{MnS}$\,(1.1)\\ $\mbox{CoO}$\,(1.6), $\mbox{VOBr}_{2}$\,(4.5)} \\
\hline\hline
\end{tabular}
\end{table*}

\textit{Discussion and summary}.
It is necessary to distinguish the odd-parity chiral magnons identified in this work from previous studies, as it has typically been associated with noncollinear antiferromagnetic order and the concomitant breaking of spin conservation\,\cite{song2025electrical, priessnitz2026ferroelectric, neumann2026odd, LuoOdd, kravchuk2026nonlinear}. In contrast, we demonstrate that the AHF phase driven by 
intra-sublattice DMI can induce odd-parity chiral magnons even in collinear AFMs, giving rise to higher-order $f$-wave and $p$-wave band splitting while preserving spin conservation. This eliminates additional spin-relaxation channels, thereby favoring long magnon lifetimes and coherent spin dynamics\,\cite{pirro2021advances, hortensius2021coherent, chen2025observation}. 

Odd-parity chiral magnon splitting enables intriguing magnon transport 
phenomena. Easy-axis $f$-wave chiral magnons generate an anomalous Hall response, while nodeless odd-parity chiral magnons enable EdE-driven spin accumulation, both of which exhibit ferroelectric switchability. As charge-neutral quasi-particles, magnons can propagate through insulators without charge carriers. Ferroelectric polarization switching can therefore reverse the sign of magnon transport without relying on a Fermi surface, avoiding the polarization instability associated with carrier accumulation that limits electronic band-splitting schemes\,\cite{duan2025antiferroelectric, Tan2019, Gumingqiang025,zhao2026multiferroic,sun2026altermagnetic,zhu2025two, wang2025two, Dong2025, fg6w-c6fd, PhysRevLett.133.096802, v3fg-6smc}. The absence of charge accumulation avoids the Joule heating, offering a route towards non-volatile and low-dissipation devices with strong ME coupling.

In summary, we uncover a symmetry-governed mechanism for realizing spin-conserving odd-parity chiral magnons in collinear antiferromagnetic multiferroics. Combining symmetry analysis with LSW calculations, we establish a unified classification of chiral splittings into $f$-wave, $p$-wave, and nodeless forms, all of which are ferroelectrically switchable. DFT calculations identify diverse bulk and 2D candidates, with chiral magnon splittings exceeding $10\,\mathrm{meV}$ in bulk multiferroics $\mbox{CoSe}$. These findings establish a robust mechanism for realizing long-lived odd-parity chiral magnons in collinear AFMs, and open new avenues for ME coupling in charge-neutral quasi-particles.

\textit{Note added}. Recently, we became aware of two relevant studies by R.\,Eto et al.\,\cite{eto2026odd} and P.\,Zhang et al.\,\cite{zhang2026odd}, whose findings partially overlap with ours.

\textit{Acknowledgment}.
The authors thank Prof.\,Bo Li for helpful discussions. This work is supported by the National Natural Science Foundation of China (Grant No.\,12374092 and No.\,T2425029), Natural Science Basic Research Program of Shaanxi (Program No.\,2023-JC-YB-017), Shaanxi Fundamental Science Research Project for Mathematics and Physics (Grant No.\,22JSQ013), “Young Talent Support Plan” of Xi'an Jiaotong University, and the Xiaomi Young Talents Program. Y.\,J.\,Jin acknowledges the National Natural Science Foundation of China (Grant No. 12404181), the Guangdong Provincial Quantum Science Strategic Initiative (Grant No. GDZX2401002).

\appendix


\bibliography{main_ref}

\begin{thebibliography}{80}%
\makeatletter
\providecommand \@ifxundefined [1]{%
 \@ifx{#1\undefined}
}%
\providecommand \@ifnum [1]{%
 \ifnum #1\expandafter \@firstoftwo
 \else \expandafter \@secondoftwo
 \fi
}%
\providecommand \@ifx [1]{%
 \ifx #1\expandafter \@firstoftwo
 \else \expandafter \@secondoftwo
 \fi
}%
\providecommand \natexlab [1]{#1}%
\providecommand \enquote  [1]{``#1''}%
\providecommand \bibnamefont  [1]{#1}%
\providecommand \bibfnamefont [1]{#1}%
\providecommand \citenamefont [1]{#1}%
\providecommand \href@noop [0]{\@secondoftwo}%
\providecommand \href [0]{\begingroup \@sanitize@url \@href}%
\providecommand \@href[1]{\@@startlink{#1}\@@href}%
\providecommand \@@href[1]{\endgroup#1\@@endlink}%
\providecommand \@sanitize@url [0]{\catcode `\\12\catcode `\$12\catcode `\&12\catcode `\#12\catcode `\^12\catcode `\_12\catcode `\%12\relax}%
\providecommand \@@startlink[1]{}%
\providecommand \@@endlink[0]{}%
\providecommand \url  [0]{\begingroup\@sanitize@url \@url }%
\providecommand \@url [1]{\endgroup\@href {#1}{\urlprefix }}%
\providecommand \urlprefix  [0]{URL }%
\providecommand \Eprint [0]{\href }%
\providecommand \doibase [0]{https://doi.org/}%
\providecommand \selectlanguage [0]{\@gobble}%
\providecommand \bibinfo  [0]{\@secondoftwo}%
\providecommand \bibfield  [0]{\@secondoftwo}%
\providecommand \translation [1]{[#1]}%
\providecommand \BibitemOpen [0]{}%
\providecommand \bibitemStop [0]{}%
\providecommand \bibitemNoStop [0]{.\EOS\space}%
\providecommand \EOS [0]{\spacefactor3000\relax}%
\providecommand \BibitemShut  [1]{\csname bibitem#1\endcsname}%
\let\auto@bib@innerbib\@empty
\bibitem [{\citenamefont {Chumak}\ \emph {et~al.}(2014)\citenamefont {Chumak}, \citenamefont {Serga},\ and\ \citenamefont {Hillebrands}}]{chumak2014magnon}%
  \BibitemOpen
  \bibfield  {author} {\bibinfo {author} {\bibfnamefont {A.~V.}\ \bibnamefont {Chumak}}, \bibinfo {author} {\bibfnamefont {A.~A.}\ \bibnamefont {Serga}},\ and\ \bibinfo {author} {\bibfnamefont {B.}~\bibnamefont {Hillebrands}},\ }\bibfield  {title} {\bibinfo {title} {Magnon transistor for all-magnon data processing},\ }\href {https://www.nature.com/articles/ncomms5700} {\bibfield  {journal} {\bibinfo  {journal} {Nat. Commun.}\ }\textbf {\bibinfo {volume} {5}},\ \bibinfo {pages} {4700} (\bibinfo {year} {2014})}\BibitemShut {NoStop}%
\bibitem [{\citenamefont {Chumak}\ \emph {et~al.}(2015)\citenamefont {Chumak}, \citenamefont {Vasyuchka}, \citenamefont {Serga},\ and\ \citenamefont {Hillebrands}}]{chumak2015magnon}%
  \BibitemOpen
  \bibfield  {author} {\bibinfo {author} {\bibfnamefont {A.~V.}\ \bibnamefont {Chumak}}, \bibinfo {author} {\bibfnamefont {V.~I.}\ \bibnamefont {Vasyuchka}}, \bibinfo {author} {\bibfnamefont {A.~A.}\ \bibnamefont {Serga}},\ and\ \bibinfo {author} {\bibfnamefont {B.}~\bibnamefont {Hillebrands}},\ }\bibfield  {title} {\bibinfo {title} {Magnon spintronics},\ }\href {https://www.nature.com/articles/nphys3347} {\bibfield  {journal} {\bibinfo  {journal} {Nat. Phys.}\ }\textbf {\bibinfo {volume} {11}},\ \bibinfo {pages} {453} (\bibinfo {year} {2015})}\BibitemShut {NoStop}%
\bibitem [{\citenamefont {Onose}\ \emph {et~al.}(2010)\citenamefont {Onose}, \citenamefont {Ideue}, \citenamefont {Katsura}, \citenamefont {Shiomi}, \citenamefont {Nagaosa},\ and\ \citenamefont {Tokura}}]{onose2010observation}%
  \BibitemOpen
  \bibfield  {author} {\bibinfo {author} {\bibfnamefont {Y.}~\bibnamefont {Onose}}, \bibinfo {author} {\bibfnamefont {T.}~\bibnamefont {Ideue}}, \bibinfo {author} {\bibfnamefont {H.}~\bibnamefont {Katsura}}, \bibinfo {author} {\bibfnamefont {Y.}~\bibnamefont {Shiomi}}, \bibinfo {author} {\bibfnamefont {N.}~\bibnamefont {Nagaosa}},\ and\ \bibinfo {author} {\bibfnamefont {Y.}~\bibnamefont {Tokura}},\ }\bibfield  {title} {\bibinfo {title} {Observation of the magnon $\mbox{Hall}$ effect},\ }\href {https://www.science.org/doi/abs/10.1126/science.1188260} {\bibfield  {journal} {\bibinfo  {journal} {Science}\ }\textbf {\bibinfo {volume} {329}},\ \bibinfo {pages} {297} (\bibinfo {year} {2010})}\BibitemShut {NoStop}%
\bibitem [{\citenamefont {Bader}\ and\ \citenamefont {Parkin}(2010)}]{bader2010spintronics}%
  \BibitemOpen
  \bibfield  {author} {\bibinfo {author} {\bibfnamefont {S.~D.}\ \bibnamefont {Bader}}\ and\ \bibinfo {author} {\bibfnamefont {S.~S.~P.}\ \bibnamefont {Parkin}},\ }\bibfield  {title} {\bibinfo {title} {Spintronics},\ }\href {https://doi.org/10.1146/annurev-conmatphys-070909-104123} {\bibfield  {journal} {\bibinfo  {journal} {Annu. Rev. Condens. Matter Phys.}\ }\textbf {\bibinfo {volume} {1}},\ \bibinfo {pages} {71} (\bibinfo {year} {2010})}\BibitemShut {NoStop}%
\bibitem [{\citenamefont {McClarty}(2022)}]{mcclarty2022topological}%
  \BibitemOpen
  \bibfield  {author} {\bibinfo {author} {\bibfnamefont {P.~A.}\ \bibnamefont {McClarty}},\ }\bibfield  {title} {\bibinfo {title} {Topological magnons: A review},\ }\href {https://doi.org/10.1146/annurev-conmatphys-031620-104715} {\bibfield  {journal} {\bibinfo  {journal} {Annu. Rev. Condens. Matter Phys.}\ }\textbf {\bibinfo {volume} {13}},\ \bibinfo {pages} {171} (\bibinfo {year} {2022})}\BibitemShut {NoStop}%
\bibitem [{\citenamefont {Lenk}\ \emph {et~al.}(2011)\citenamefont {Lenk}, \citenamefont {Ulrichs}, \citenamefont {Garbs},\ and\ \citenamefont {M{\"u}nzenberg}}]{lenk2011building}%
  \BibitemOpen
  \bibfield  {author} {\bibinfo {author} {\bibfnamefont {B.}~\bibnamefont {Lenk}}, \bibinfo {author} {\bibfnamefont {H.}~\bibnamefont {Ulrichs}}, \bibinfo {author} {\bibfnamefont {F.}~\bibnamefont {Garbs}},\ and\ \bibinfo {author} {\bibfnamefont {M.}~\bibnamefont {M{\"u}nzenberg}},\ }\bibfield  {title} {\bibinfo {title} {The building blocks of magnonics},\ }\href {https://www.sciencedirect.com/science/article/abs/pii/S0370157311001694} {\bibfield  {journal} {\bibinfo  {journal} {Phys. Rep.}\ }\textbf {\bibinfo {volume} {507}},\ \bibinfo {pages} {107} (\bibinfo {year} {2011})}\BibitemShut {NoStop}%
\bibitem [{\citenamefont {Chen}\ \emph {et~al.}(2025{\natexlab{a}})\citenamefont {Chen}, \citenamefont {Liu}, \citenamefont {Liu}, \citenamefont {Yu}, \citenamefont {Ren}, \citenamefont {Li}, \citenamefont {Zhang},\ and\ \citenamefont {Liu}}]{chen2025unconventional}%
  \BibitemOpen
  \bibfield  {author} {\bibinfo {author} {\bibfnamefont {X.}~\bibnamefont {Chen}}, \bibinfo {author} {\bibfnamefont {Y.}~\bibnamefont {Liu}}, \bibinfo {author} {\bibfnamefont {P.}~\bibnamefont {Liu}}, \bibinfo {author} {\bibfnamefont {Y.}~\bibnamefont {Yu}}, \bibinfo {author} {\bibfnamefont {J.}~\bibnamefont {Ren}}, \bibinfo {author} {\bibfnamefont {J.}~\bibnamefont {Li}}, \bibinfo {author} {\bibfnamefont {A.}~\bibnamefont {Zhang}},\ and\ \bibinfo {author} {\bibfnamefont {Q.}~\bibnamefont {Liu}},\ }\bibfield  {title} {\bibinfo {title} {Unconventional magnons in collinear magnets dictated by spin space groups},\ }\href {https://www.nature.com/articles/s41586-025-08715-7} {\bibfield  {journal} {\bibinfo  {journal} {Nature}\ }\textbf {\bibinfo {volume} {640}},\ \bibinfo {pages} {349} (\bibinfo {year} {2025}{\natexlab{a}})}\BibitemShut {NoStop}%
\bibitem [{\citenamefont {Chen}\ \emph {et~al.}(2018)\citenamefont {Chen}, \citenamefont {Chung}, \citenamefont {Gao}, \citenamefont {Chen}, \citenamefont {Stone}, \citenamefont {Kolesnikov}, \citenamefont {Huang},\ and\ \citenamefont {Dai}}]{chen_PRX_2018_8}%
  \BibitemOpen
  \bibfield  {author} {\bibinfo {author} {\bibfnamefont {L.}~\bibnamefont {Chen}}, \bibinfo {author} {\bibfnamefont {J.-H.}\ \bibnamefont {Chung}}, \bibinfo {author} {\bibfnamefont {B.}~\bibnamefont {Gao}}, \bibinfo {author} {\bibfnamefont {T.}~\bibnamefont {Chen}}, \bibinfo {author} {\bibfnamefont {M.~B.}\ \bibnamefont {Stone}}, \bibinfo {author} {\bibfnamefont {A.~I.}\ \bibnamefont {Kolesnikov}}, \bibinfo {author} {\bibfnamefont {Q.}~\bibnamefont {Huang}},\ and\ \bibinfo {author} {\bibfnamefont {P.}~\bibnamefont {Dai}},\ }\bibfield  {title} {\bibinfo {title} {Topological spin excitations in honeycomb ferromagnet $\mbox{CrI}_{3}$},\ }\href {https://journals.aps.org/prx/abstract/10.1103/PhysRevX.8.041028} {\bibfield  {journal} {\bibinfo  {journal} {Phys. Rev. X}\ }\textbf {\bibinfo {volume} {8}},\ \bibinfo {pages} {041028} (\bibinfo {year} {2018})}\BibitemShut {NoStop}%
\bibitem [{\citenamefont {Mook}\ \emph {et~al.}(2021)\citenamefont {Mook}, \citenamefont {Plekhanov}, \citenamefont {Klinovaja},\ and\ \citenamefont {Loss}}]{mook_PRX_2021_11}%
  \BibitemOpen
  \bibfield  {author} {\bibinfo {author} {\bibfnamefont {A.}~\bibnamefont {Mook}}, \bibinfo {author} {\bibfnamefont {K.}~\bibnamefont {Plekhanov}}, \bibinfo {author} {\bibfnamefont {J.}~\bibnamefont {Klinovaja}},\ and\ \bibinfo {author} {\bibfnamefont {D.}~\bibnamefont {Loss}},\ }\bibfield  {title} {\bibinfo {title} {Interaction-stabilized topological magnon insulator in ferromagnets},\ }\href {https://journals.aps.org/prx/abstract/10.1103/PhysRevX.11.021061} {\bibfield  {journal} {\bibinfo  {journal} {Phys. Rev. X}\ }\textbf {\bibinfo {volume} {11}},\ \bibinfo {pages} {021061} (\bibinfo {year} {2021})}\BibitemShut {NoStop}%
\bibitem [{\citenamefont {Wang}\ \emph {et~al.}(2019)\citenamefont {Wang}, \citenamefont {Rao}, \citenamefont {Yang}, \citenamefont {Xu}, \citenamefont {Gui}, \citenamefont {Yao}, \citenamefont {You},\ and\ \citenamefont {Hu}}]{wang2019nonreciprocity}%
  \BibitemOpen
  \bibfield  {author} {\bibinfo {author} {\bibfnamefont {Y.-P.}\ \bibnamefont {Wang}}, \bibinfo {author} {\bibfnamefont {J.~W.}\ \bibnamefont {Rao}}, \bibinfo {author} {\bibfnamefont {Y.}~\bibnamefont {Yang}}, \bibinfo {author} {\bibfnamefont {P.-C.}\ \bibnamefont {Xu}}, \bibinfo {author} {\bibfnamefont {Y.~S.}\ \bibnamefont {Gui}}, \bibinfo {author} {\bibfnamefont {B.~M.}\ \bibnamefont {Yao}}, \bibinfo {author} {\bibfnamefont {J.~Q.}\ \bibnamefont {You}},\ and\ \bibinfo {author} {\bibfnamefont {C.-M.}\ \bibnamefont {Hu}},\ }\bibfield  {title} {\bibinfo {title} {Nonreciprocity and unidirectional invisibility in cavity magnonics},\ }\href {https://link.aps.org/doi/10.1103/PhysRevLett.123.127202} {\bibfield  {journal} {\bibinfo  {journal} {Phys. Rev. Lett.}\ }\textbf {\bibinfo {volume} {123}},\ \bibinfo {pages} {127202} (\bibinfo {year} {2019})}\BibitemShut {NoStop}%
\bibitem [{\citenamefont {Bai}\ \emph {et~al.}(2026)\citenamefont {Bai}, \citenamefont {Yuan}, \citenamefont {Chen}, \citenamefont {Dai}, \citenamefont {Huang}, \citenamefont {Wang},\ and\ \citenamefont {Niu}}]{bai2026ferroelectrics}%
  \BibitemOpen
  \bibfield  {author} {\bibinfo {author} {\bibfnamefont {Y.}~\bibnamefont {Bai}}, \bibinfo {author} {\bibfnamefont {B.}~\bibnamefont {Yuan}}, \bibinfo {author} {\bibfnamefont {Z.}~\bibnamefont {Chen}}, \bibinfo {author} {\bibfnamefont {Y.}~\bibnamefont {Dai}}, \bibinfo {author} {\bibfnamefont {B.}~\bibnamefont {Huang}}, \bibinfo {author} {\bibfnamefont {X.}~\bibnamefont {Wang}},\ and\ \bibinfo {author} {\bibfnamefont {C.}~\bibnamefont {Niu}},\ }\bibfield  {title} {\bibinfo {title} {Ferroelectrics drive topological magnon transitions and valley transport},\ }\href {https://journals.aps.org/prl/abstract/10.1103/588c-z8cy} {\bibfield  {journal} {\bibinfo  {journal} {Phys. Rev. Lett.}\ }\textbf {\bibinfo {volume} {136}},\ \bibinfo {pages} {046602} (\bibinfo {year} {2026})}\BibitemShut {NoStop}%
\bibitem [{\citenamefont {Baltz}\ \emph {et~al.}(2018)\citenamefont {Baltz}, \citenamefont {Manchon}, \citenamefont {Tsoi}, \citenamefont {Moriyama}, \citenamefont {Ono},\ and\ \citenamefont {Tserkovnyak}}]{baltz2018antiferromagnetic}%
  \BibitemOpen
  \bibfield  {author} {\bibinfo {author} {\bibfnamefont {V.}~\bibnamefont {Baltz}}, \bibinfo {author} {\bibfnamefont {A.}~\bibnamefont {Manchon}}, \bibinfo {author} {\bibfnamefont {M.}~\bibnamefont {Tsoi}}, \bibinfo {author} {\bibfnamefont {T.}~\bibnamefont {Moriyama}}, \bibinfo {author} {\bibfnamefont {T.}~\bibnamefont {Ono}},\ and\ \bibinfo {author} {\bibfnamefont {Y.}~\bibnamefont {Tserkovnyak}},\ }\bibfield  {title} {\bibinfo {title} {Antiferromagnetic spintronics},\ }\href {https://journals.aps.org/rmp/abstract/10.1103/RevModPhys.90.015005} {\bibfield  {journal} {\bibinfo  {journal} {Rev. Mod. Phys.}\ }\textbf {\bibinfo {volume} {90}},\ \bibinfo {pages} {015005} (\bibinfo {year} {2018})}\BibitemShut {NoStop}%
\bibitem [{\citenamefont {Jungwirth}\ \emph {et~al.}(2016)\citenamefont {Jungwirth}, \citenamefont {Marti}, \citenamefont {Wadley},\ and\ \citenamefont {Wunderlich}}]{jungwirth2016antiferromagnetic}%
  \BibitemOpen
  \bibfield  {author} {\bibinfo {author} {\bibfnamefont {T.}~\bibnamefont {Jungwirth}}, \bibinfo {author} {\bibfnamefont {X.}~\bibnamefont {Marti}}, \bibinfo {author} {\bibfnamefont {P.}~\bibnamefont {Wadley}},\ and\ \bibinfo {author} {\bibfnamefont {J.}~\bibnamefont {Wunderlich}},\ }\bibfield  {title} {\bibinfo {title} {Antiferromagnetic spintronics},\ }\href {https://www.nature.com/articles/nnano.2016.18} {\bibfield  {journal} {\bibinfo  {journal} {Nat. Nanotechnol.}\ }\textbf {\bibinfo {volume} {11}},\ \bibinfo {pages} {231} (\bibinfo {year} {2016})}\BibitemShut {NoStop}%
\bibitem [{\citenamefont {Rezende}\ \emph {et~al.}(2019)\citenamefont {Rezende}, \citenamefont {Azevedo},\ and\ \citenamefont {Rodr{\'\i}guez-Su{\'a}rez}}]{rezende2019introduction}%
  \BibitemOpen
  \bibfield  {author} {\bibinfo {author} {\bibfnamefont {S.~M.}\ \bibnamefont {Rezende}}, \bibinfo {author} {\bibfnamefont {A.}~\bibnamefont {Azevedo}},\ and\ \bibinfo {author} {\bibfnamefont {R.~L.}\ \bibnamefont {Rodr{\'\i}guez-Su{\'a}rez}},\ }\bibfield  {title} {\bibinfo {title} {Introduction to antiferromagnetic magnons},\ }\href {https://pubs.aip.org/aip/jap/article/126/15/151101/994488} {\bibfield  {journal} {\bibinfo  {journal} {J. Appl. Phys.}\ }\textbf {\bibinfo {volume} {126}},\ \bibinfo {pages} {151101} (\bibinfo {year} {2019})}\BibitemShut {NoStop}%
\bibitem [{\citenamefont {{\v{S}}mejkal}\ \emph {et~al.}(2020)\citenamefont {{\v{S}}mejkal}, \citenamefont {Gonz{\'a}lez-Hern{\'a}ndez}, \citenamefont {Jungwirth},\ and\ \citenamefont {Sinova}}]{vsmejkal2020crystal}%
  \BibitemOpen
  \bibfield  {author} {\bibinfo {author} {\bibfnamefont {L.}~\bibnamefont {{\v{S}}mejkal}}, \bibinfo {author} {\bibfnamefont {R.}~\bibnamefont {Gonz{\'a}lez-Hern{\'a}ndez}}, \bibinfo {author} {\bibfnamefont {T.}~\bibnamefont {Jungwirth}},\ and\ \bibinfo {author} {\bibfnamefont {J.}~\bibnamefont {Sinova}},\ }\bibfield  {title} {\bibinfo {title} {Crystal time-reversal symmetry breaking and spontaneous $\mbox{Hall}$ effect in collinear antiferromagnets},\ }\href {https://www.science.org/doi/full/10.1126/sciadv.aaz8809} {\bibfield  {journal} {\bibinfo  {journal} {Sci. Adv.}\ }\textbf {\bibinfo {volume} {6}},\ \bibinfo {pages} {eaaz8809} (\bibinfo {year} {2020})}\BibitemShut {NoStop}%
\bibitem [{\citenamefont {Mazin}\ \emph {et~al.}(2021)\citenamefont {Mazin}, \citenamefont {Koepernik}, \citenamefont {Johannes}, \citenamefont {Gonz{\'a}lez-Hern{\'a}ndez},\ and\ \citenamefont {{\v{S}}mejkal}}]{mazin2021prediction}%
  \BibitemOpen
  \bibfield  {author} {\bibinfo {author} {\bibfnamefont {I.~I.}\ \bibnamefont {Mazin}}, \bibinfo {author} {\bibfnamefont {K.}~\bibnamefont {Koepernik}}, \bibinfo {author} {\bibfnamefont {M.~D.}\ \bibnamefont {Johannes}}, \bibinfo {author} {\bibfnamefont {R.}~\bibnamefont {Gonz{\'a}lez-Hern{\'a}ndez}},\ and\ \bibinfo {author} {\bibfnamefont {L.}~\bibnamefont {{\v{S}}mejkal}},\ }\bibfield  {title} {\bibinfo {title} {Prediction of unconventional magnetism in doped $\mbox{FeSb}_{2}$},\ }\href {https://www.pnas.org/doi/abs/10.1073/pnas.2108924118} {\bibfield  {journal} {\bibinfo  {journal} {Proc. Natl. Acad. Sci.}\ }\textbf {\bibinfo {volume} {118}},\ \bibinfo {pages} {e2108924118} (\bibinfo {year} {2021})}\BibitemShut {NoStop}%
\bibitem [{\citenamefont {Bai}\ \emph {et~al.}(2024)\citenamefont {Bai}, \citenamefont {Feng}, \citenamefont {Liu}, \citenamefont {{\v{S}}mejkal}, \citenamefont {Mokrousov},\ and\ \citenamefont {Yao}}]{bai2024altermagnetism}%
  \BibitemOpen
  \bibfield  {author} {\bibinfo {author} {\bibfnamefont {L.}~\bibnamefont {Bai}}, \bibinfo {author} {\bibfnamefont {W.}~\bibnamefont {Feng}}, \bibinfo {author} {\bibfnamefont {S.}~\bibnamefont {Liu}}, \bibinfo {author} {\bibfnamefont {L.}~\bibnamefont {{\v{S}}mejkal}}, \bibinfo {author} {\bibfnamefont {Y.}~\bibnamefont {Mokrousov}},\ and\ \bibinfo {author} {\bibfnamefont {Y.}~\bibnamefont {Yao}},\ }\bibfield  {title} {\bibinfo {title} {Altermagnetism: Exploring new frontiers in magnetism and spintronics},\ }\href {https://advanced.onlinelibrary.wiley.com/doi/abs/10.1002/adfm.202409327} {\bibfield  {journal} {\bibinfo  {journal} {Adv. Funct. Mater.}\ }\textbf {\bibinfo {volume} {34}},\ \bibinfo {pages} {2409327} (\bibinfo {year} {2024})}\BibitemShut {NoStop}%
\bibitem [{\citenamefont {Zhou}\ \emph {et~al.}(2025{\natexlab{a}})\citenamefont {Zhou}, \citenamefont {Cheng}, \citenamefont {Hu}, \citenamefont {Chu}, \citenamefont {Bai}, \citenamefont {Han}, \citenamefont {Liu}, \citenamefont {Pan},\ and\ \citenamefont {Song}}]{zhou2025manipulation}%
  \BibitemOpen
  \bibfield  {author} {\bibinfo {author} {\bibfnamefont {Z.}~\bibnamefont {Zhou}}, \bibinfo {author} {\bibfnamefont {X.}~\bibnamefont {Cheng}}, \bibinfo {author} {\bibfnamefont {M.}~\bibnamefont {Hu}}, \bibinfo {author} {\bibfnamefont {R.}~\bibnamefont {Chu}}, \bibinfo {author} {\bibfnamefont {H.}~\bibnamefont {Bai}}, \bibinfo {author} {\bibfnamefont {L.}~\bibnamefont {Han}}, \bibinfo {author} {\bibfnamefont {J.}~\bibnamefont {Liu}}, \bibinfo {author} {\bibfnamefont {F.}~\bibnamefont {Pan}},\ and\ \bibinfo {author} {\bibfnamefont {C.}~\bibnamefont {Song}},\ }\bibfield  {title} {\bibinfo {title} {Manipulation of the altermagnetic order in $\mbox{CrSb}$ via crystal symmetry},\ }\href {https://www.nature.com/articles/s41586-024-08436-3} {\bibfield  {journal} {\bibinfo  {journal} {Nature}\ }\textbf {\bibinfo {volume} {638}},\ \bibinfo {pages} {645} (\bibinfo {year} {2025}{\natexlab{a}})}\BibitemShut {NoStop}%
\bibitem [{\citenamefont {Duan}\ \emph {et~al.}(2025)\citenamefont {Duan}, \citenamefont {Zhang}, \citenamefont {Zhu}, \citenamefont {Liu}, \citenamefont {Zhang}, \citenamefont {{\v{Z}}uti{\'c}},\ and\ \citenamefont {Zhou}}]{duan2025antiferroelectric}%
  \BibitemOpen
  \bibfield  {author} {\bibinfo {author} {\bibfnamefont {X.}~\bibnamefont {Duan}}, \bibinfo {author} {\bibfnamefont {J.}~\bibnamefont {Zhang}}, \bibinfo {author} {\bibfnamefont {Z.}~\bibnamefont {Zhu}}, \bibinfo {author} {\bibfnamefont {Y.}~\bibnamefont {Liu}}, \bibinfo {author} {\bibfnamefont {Z.}~\bibnamefont {Zhang}}, \bibinfo {author} {\bibfnamefont {I.}~\bibnamefont {{\v{Z}}uti{\'c}}},\ and\ \bibinfo {author} {\bibfnamefont {T.}~\bibnamefont {Zhou}},\ }\bibfield  {title} {\bibinfo {title} {Antiferroelectric altermagnets: Antiferroelectricity alters magnets},\ }\href {https://journals.aps.org/prl/abstract/10.1103/PhysRevLett.134.106801} {\bibfield  {journal} {\bibinfo  {journal} {Phys. Rev. Lett.}\ }\textbf {\bibinfo {volume} {134}},\ \bibinfo {pages} {106801} (\bibinfo {year} {2025})}\BibitemShut {NoStop}%
\bibitem [{\citenamefont {Song}\ \emph {et~al.}(2025{\natexlab{a}})\citenamefont {Song}, \citenamefont {Bai}, \citenamefont {Zhou}, \citenamefont {Han}, \citenamefont {Reichlova}, \citenamefont {Dil}, \citenamefont {Liu}, \citenamefont {Chen},\ and\ \citenamefont {Pan}}]{song2025altermagnets}%
  \BibitemOpen
  \bibfield  {author} {\bibinfo {author} {\bibfnamefont {C.}~\bibnamefont {Song}}, \bibinfo {author} {\bibfnamefont {H.}~\bibnamefont {Bai}}, \bibinfo {author} {\bibfnamefont {Z.}~\bibnamefont {Zhou}}, \bibinfo {author} {\bibfnamefont {L.}~\bibnamefont {Han}}, \bibinfo {author} {\bibfnamefont {H.}~\bibnamefont {Reichlova}}, \bibinfo {author} {\bibfnamefont {J.~H.}\ \bibnamefont {Dil}}, \bibinfo {author} {\bibfnamefont {J.}~\bibnamefont {Liu}}, \bibinfo {author} {\bibfnamefont {X.}~\bibnamefont {Chen}},\ and\ \bibinfo {author} {\bibfnamefont {F.}~\bibnamefont {Pan}},\ }\bibfield  {title} {\bibinfo {title} {Altermagnets as a new class of functional materials},\ }\href {https://www.nature.com/articles/s41578-025-00779-1} {\bibfield  {journal} {\bibinfo  {journal} {Nat. Rev. Mater.}\ }\textbf {\bibinfo {volume} {10}},\ \bibinfo {pages} {473} (\bibinfo {year} {2025}{\natexlab{a}})}\BibitemShut {NoStop}%
\bibitem [{\citenamefont {Liu}\ \emph {et~al.}(2022)\citenamefont {Liu}, \citenamefont {Li}, \citenamefont {Han}, \citenamefont {Wan},\ and\ \citenamefont {Liu}}]{liu2022spin}%
  \BibitemOpen
  \bibfield  {author} {\bibinfo {author} {\bibfnamefont {P.}~\bibnamefont {Liu}}, \bibinfo {author} {\bibfnamefont {J.}~\bibnamefont {Li}}, \bibinfo {author} {\bibfnamefont {J.}~\bibnamefont {Han}}, \bibinfo {author} {\bibfnamefont {X.}~\bibnamefont {Wan}},\ and\ \bibinfo {author} {\bibfnamefont {Q.}~\bibnamefont {Liu}},\ }\bibfield  {title} {\bibinfo {title} {Spin-group symmetry in magnetic materials with negligible spin-orbit coupling},\ }\href {https://journals.aps.org/prx/abstract/10.1103/PhysRevX.12.021016} {\bibfield  {journal} {\bibinfo  {journal} {Phys. Rev. X}\ }\textbf {\bibinfo {volume} {12}},\ \bibinfo {pages} {021016} (\bibinfo {year} {2022})}\BibitemShut {NoStop}%
\bibitem [{\citenamefont {\v{S}mejkal}\ \emph {et~al.}(2023)\citenamefont {\v{S}mejkal}, \citenamefont {Marmodoro}, \citenamefont {Ahn}, \citenamefont {Gonz\'alez-Hern\'andez}, \citenamefont {Turek}, \citenamefont {Mankovsky}, \citenamefont {Ebert}, \citenamefont {D'Souza}, \citenamefont {{\v{S}}ipr}, \citenamefont {Sinova},\ and\ \citenamefont {Jungwirth}}]{vsmejkal2023chiral}%
  \BibitemOpen
  \bibfield  {author} {\bibinfo {author} {\bibfnamefont {L.}~\bibnamefont {\v{S}mejkal}}, \bibinfo {author} {\bibfnamefont {A.}~\bibnamefont {Marmodoro}}, \bibinfo {author} {\bibfnamefont {K.-H.}\ \bibnamefont {Ahn}}, \bibinfo {author} {\bibfnamefont {R.}~\bibnamefont {Gonz\'alez-Hern\'andez}}, \bibinfo {author} {\bibfnamefont {I.}~\bibnamefont {Turek}}, \bibinfo {author} {\bibfnamefont {S.}~\bibnamefont {Mankovsky}}, \bibinfo {author} {\bibfnamefont {H.}~\bibnamefont {Ebert}}, \bibinfo {author} {\bibfnamefont {S.~W.}\ \bibnamefont {D'Souza}}, \bibinfo {author} {\bibfnamefont {O.}~\bibnamefont {{\v{S}}ipr}}, \bibinfo {author} {\bibfnamefont {J.}~\bibnamefont {Sinova}},\ and\ \bibinfo {author} {\bibfnamefont {T.}~\bibnamefont {Jungwirth}},\ }\bibfield  {title} {\bibinfo {title} {Chiral magnons in altermagnetic $\mbox{RuO}_{2}$},\ }\href {https://journals.aps.org/prl/abstract/10.1103/PhysRevLett.131.256703} {\bibfield  {journal} {\bibinfo  {journal} {Phys. Rev. Lett.}\ }\textbf {\bibinfo {volume} {131}},\
  \bibinfo {pages} {256703} (\bibinfo {year} {2023})}\BibitemShut {NoStop}%
\bibitem [{\citenamefont {Hoyer}\ \emph {et~al.}(2025)\citenamefont {Hoyer}, \citenamefont {Stavropoulos}, \citenamefont {Razpopov}, \citenamefont {Valent{\'\i}}, \citenamefont {{\v{S}}mejkal},\ and\ \citenamefont {Mook}}]{hoyer2025altermagnetic}%
  \BibitemOpen
  \bibfield  {author} {\bibinfo {author} {\bibfnamefont {R.}~\bibnamefont {Hoyer}}, \bibinfo {author} {\bibfnamefont {P.~P.}\ \bibnamefont {Stavropoulos}}, \bibinfo {author} {\bibfnamefont {A.}~\bibnamefont {Razpopov}}, \bibinfo {author} {\bibfnamefont {R.}~\bibnamefont {Valent{\'\i}}}, \bibinfo {author} {\bibfnamefont {L.}~\bibnamefont {{\v{S}}mejkal}},\ and\ \bibinfo {author} {\bibfnamefont {A.}~\bibnamefont {Mook}},\ }\bibfield  {title} {\bibinfo {title} {Altermagnetic splitting of magnons in hematite $\alpha$-$\mbox{Fe}_{2}\mbox{O}_{3}$},\ }\href {https://journals.aps.org/prb/abstract/10.1103/fgc1-5blp} {\bibfield  {journal} {\bibinfo  {journal} {Phys. Rev. B}\ }\textbf {\bibinfo {volume} {112}},\ \bibinfo {pages} {064425} (\bibinfo {year} {2025})}\BibitemShut {NoStop}%
\bibitem [{\citenamefont {Cui}\ \emph {et~al.}(2023)\citenamefont {Cui}, \citenamefont {Zeng}, \citenamefont {Cui}, \citenamefont {Yu},\ and\ \citenamefont {Yang}}]{cui2023efficient}%
  \BibitemOpen
  \bibfield  {author} {\bibinfo {author} {\bibfnamefont {Q.}~\bibnamefont {Cui}}, \bibinfo {author} {\bibfnamefont {B.}~\bibnamefont {Zeng}}, \bibinfo {author} {\bibfnamefont {P.}~\bibnamefont {Cui}}, \bibinfo {author} {\bibfnamefont {T.}~\bibnamefont {Yu}},\ and\ \bibinfo {author} {\bibfnamefont {H.}~\bibnamefont {Yang}},\ }\bibfield  {title} {\bibinfo {title} {Efficient spin $\mbox{Seebeck}$ and spin $\mbox{Nernst}$ effects of magnons in altermagnets},\ }\href {https://journals.aps.org/prb/abstract/10.1103/PhysRevB.108.L180401} {\bibfield  {journal} {\bibinfo  {journal} {Phys. Rev. B}\ }\textbf {\bibinfo {volume} {108}},\ \bibinfo {pages} {L180401} (\bibinfo {year} {2023})}\BibitemShut {NoStop}%
\bibitem [{\citenamefont {Liu}\ \emph {et~al.}(2024)\citenamefont {Liu}, \citenamefont {Ozeki}, \citenamefont {Asai}, \citenamefont {Itoh},\ and\ \citenamefont {Masuda}}]{liu2024chiral}%
  \BibitemOpen
  \bibfield  {author} {\bibinfo {author} {\bibfnamefont {Z.}~\bibnamefont {Liu}}, \bibinfo {author} {\bibfnamefont {M.}~\bibnamefont {Ozeki}}, \bibinfo {author} {\bibfnamefont {S.}~\bibnamefont {Asai}}, \bibinfo {author} {\bibfnamefont {S.}~\bibnamefont {Itoh}},\ and\ \bibinfo {author} {\bibfnamefont {T.}~\bibnamefont {Masuda}},\ }\bibfield  {title} {\bibinfo {title} {Chiral split magnon in altermagnetic $\mbox{MnTe}$},\ }\href {https://journals.aps.org/prl/abstract/10.1103/PhysRevLett.133.156702} {\bibfield  {journal} {\bibinfo  {journal} {Phys. Rev. Lett.}\ }\textbf {\bibinfo {volume} {133}},\ \bibinfo {pages} {156702} (\bibinfo {year} {2024})}\BibitemShut {NoStop}%
\bibitem [{\citenamefont {Sun}\ \emph {et~al.}(2025)\citenamefont {Sun}, \citenamefont {Guo}, \citenamefont {Wang}, \citenamefont {Abernathy}, \citenamefont {Tian},\ and\ \citenamefont {Li}}]{sun2025observation}%
  \BibitemOpen
  \bibfield  {author} {\bibinfo {author} {\bibfnamefont {Q.}~\bibnamefont {Sun}}, \bibinfo {author} {\bibfnamefont {J.}~\bibnamefont {Guo}}, \bibinfo {author} {\bibfnamefont {D.}~\bibnamefont {Wang}}, \bibinfo {author} {\bibfnamefont {D.~L.}\ \bibnamefont {Abernathy}}, \bibinfo {author} {\bibfnamefont {W.}~\bibnamefont {Tian}},\ and\ \bibinfo {author} {\bibfnamefont {C.}~\bibnamefont {Li}},\ }\bibfield  {title} {\bibinfo {title} {Observation of chiral magnon band splitting in altermagnetic hematite},\ }\href {https://journals.aps.org/prl/abstract/10.1103/7yhz-jptc} {\bibfield  {journal} {\bibinfo  {journal} {Phys. Rev. Lett.}\ }\textbf {\bibinfo {volume} {135}},\ \bibinfo {pages} {186703} (\bibinfo {year} {2025})}\BibitemShut {NoStop}%
\bibitem [{\citenamefont {Singh}\ \emph {et~al.}(2025)\citenamefont {Singh}, \citenamefont {Heinsdorf}, \citenamefont {Mancilla}, \citenamefont {Bannies}, \citenamefont {Maity}, \citenamefont {Kolesnikov}, \citenamefont {Matsuda}, \citenamefont {Stone}, \citenamefont {Franz}, \citenamefont {Gaudet},\ and\ \citenamefont {Hallas}}]{singh2025chiral}%
  \BibitemOpen
  \bibfield  {author} {\bibinfo {author} {\bibfnamefont {A.~K.}\ \bibnamefont {Singh}}, \bibinfo {author} {\bibfnamefont {N.}~\bibnamefont {Heinsdorf}}, \bibinfo {author} {\bibfnamefont {A.~A.}\ \bibnamefont {Mancilla}}, \bibinfo {author} {\bibfnamefont {J.}~\bibnamefont {Bannies}}, \bibinfo {author} {\bibfnamefont {A.}~\bibnamefont {Maity}}, \bibinfo {author} {\bibfnamefont {A.~I.}\ \bibnamefont {Kolesnikov}}, \bibinfo {author} {\bibfnamefont {M.}~\bibnamefont {Matsuda}}, \bibinfo {author} {\bibfnamefont {M.~B.}\ \bibnamefont {Stone}}, \bibinfo {author} {\bibfnamefont {M.}~\bibnamefont {Franz}}, \bibinfo {author} {\bibfnamefont {J.}~\bibnamefont {Gaudet}},\ and\ \bibinfo {author} {\bibfnamefont {A.~M.}\ \bibnamefont {Hallas}},\ }\bibfield  {title} {\bibinfo {title} {Chiral spin-split magnons in the metallic altermagnet $\mbox{CrSb}$},\ }\href {https://arxiv.org/abs/2511.16086} {\bibfield  {journal} {\bibinfo  {journal} {arXiv:2511.16086}\ } (\bibinfo {year} {2025})}\BibitemShut {NoStop}%
\bibitem [{\citenamefont {Morano}\ \emph {et~al.}(2025)\citenamefont {Morano}, \citenamefont {Maesen}, \citenamefont {Nikitin}, \citenamefont {Lass}, \citenamefont {Mazzone},\ and\ \citenamefont {Zaharko}}]{morano2025absence}%
  \BibitemOpen
  \bibfield  {author} {\bibinfo {author} {\bibfnamefont {V.~C.}\ \bibnamefont {Morano}}, \bibinfo {author} {\bibfnamefont {Z.}~\bibnamefont {Maesen}}, \bibinfo {author} {\bibfnamefont {S.~E.}\ \bibnamefont {Nikitin}}, \bibinfo {author} {\bibfnamefont {J.}~\bibnamefont {Lass}}, \bibinfo {author} {\bibfnamefont {D.~G.}\ \bibnamefont {Mazzone}},\ and\ \bibinfo {author} {\bibfnamefont {O.}~\bibnamefont {Zaharko}},\ }\bibfield  {title} {\bibinfo {title} {Absence of altermagnetic magnon band splitting in $\mbox{MnF}_{2}$},\ }\href {https://journals.aps.org/prl/abstract/10.1103/PhysRevLett.134.226702} {\bibfield  {journal} {\bibinfo  {journal} {Phys. Rev. Lett.}\ }\textbf {\bibinfo {volume} {134}},\ \bibinfo {pages} {226702} (\bibinfo {year} {2025})}\BibitemShut {NoStop}%
\bibitem [{\citenamefont {Song}\ \emph {et~al.}(2025{\natexlab{b}})\citenamefont {Song}, \citenamefont {Stavri{\'c}}, \citenamefont {Barone}, \citenamefont {Droghetti}, \citenamefont {Antonenko}, \citenamefont {Venderbos}, \citenamefont {Occhialini}, \citenamefont {Ilyas}, \citenamefont {Erge{\c{c}}en}, \citenamefont {Gedik}, \citenamefont {Cheong}, \citenamefont {Fernandes}, \citenamefont {Picozzi},\ and\ \citenamefont {Comin}}]{song2025electrical}%
  \BibitemOpen
  \bibfield  {author} {\bibinfo {author} {\bibfnamefont {Q.}~\bibnamefont {Song}}, \bibinfo {author} {\bibfnamefont {S.}~\bibnamefont {Stavri{\'c}}}, \bibinfo {author} {\bibfnamefont {P.}~\bibnamefont {Barone}}, \bibinfo {author} {\bibfnamefont {A.}~\bibnamefont {Droghetti}}, \bibinfo {author} {\bibfnamefont {D.~S.}\ \bibnamefont {Antonenko}}, \bibinfo {author} {\bibfnamefont {J.~W.~F.}\ \bibnamefont {Venderbos}}, \bibinfo {author} {\bibfnamefont {C.~A.}\ \bibnamefont {Occhialini}}, \bibinfo {author} {\bibfnamefont {B.}~\bibnamefont {Ilyas}}, \bibinfo {author} {\bibfnamefont {E.}~\bibnamefont {Erge{\c{c}}en}}, \bibinfo {author} {\bibfnamefont {N.}~\bibnamefont {Gedik}}, \bibinfo {author} {\bibfnamefont {S.-W.}\ \bibnamefont {Cheong}}, \bibinfo {author} {\bibfnamefont {R.~M.}\ \bibnamefont {Fernandes}}, \bibinfo {author} {\bibfnamefont {S.}~\bibnamefont {Picozzi}},\ and\ \bibinfo {author} {\bibfnamefont {R.}~\bibnamefont {Comin}},\ }\bibfield  {title} {\bibinfo {title} {Electrical switching of a $p$-wave
  magnet},\ }\href {https://www.nature.com/articles/s41586-025-09034-7} {\bibfield  {journal} {\bibinfo  {journal} {Nature}\ }\textbf {\bibinfo {volume} {642}},\ \bibinfo {pages} {64} (\bibinfo {year} {2025}{\natexlab{b}})}\BibitemShut {NoStop}%
\bibitem [{\citenamefont {Priessnitz}\ \emph {et~al.}(2026)\citenamefont {Priessnitz}, \citenamefont {Hellenes}, \citenamefont {Comin},\ and\ \citenamefont {{\v{S}}mejkal}}]{priessnitz2026ferroelectric}%
  \BibitemOpen
  \bibfield  {author} {\bibinfo {author} {\bibfnamefont {J.}~\bibnamefont {Priessnitz}}, \bibinfo {author} {\bibfnamefont {A.~B.}\ \bibnamefont {Hellenes}}, \bibinfo {author} {\bibfnamefont {R.}~\bibnamefont {Comin}},\ and\ \bibinfo {author} {\bibfnamefont {L.}~\bibnamefont {{\v{S}}mejkal}},\ }\bibfield  {title} {\bibinfo {title} {Ferroelectric $p$-wave magnets},\ }\href {https://arxiv.org/abs/2603.19107} {\bibfield  {journal} {\bibinfo  {journal} {arXiv:2603.19107}\ } (\bibinfo {year} {2026})}\BibitemShut {NoStop}%
\bibitem [{\citenamefont {Neumann}\ \emph {et~al.}(2026)\citenamefont {Neumann}, \citenamefont {Jaeschke-Ubiergo}, \citenamefont {Zarzuela}, \citenamefont {{\v{S}}mejkal}, \citenamefont {Sinova},\ and\ \citenamefont {Mook}}]{neumann2026odd}%
  \BibitemOpen
  \bibfield  {author} {\bibinfo {author} {\bibfnamefont {R.~R.}\ \bibnamefont {Neumann}}, \bibinfo {author} {\bibfnamefont {R.}~\bibnamefont {Jaeschke-Ubiergo}}, \bibinfo {author} {\bibfnamefont {R.}~\bibnamefont {Zarzuela}}, \bibinfo {author} {\bibfnamefont {L.}~\bibnamefont {{\v{S}}mejkal}}, \bibinfo {author} {\bibfnamefont {J.}~\bibnamefont {Sinova}},\ and\ \bibinfo {author} {\bibfnamefont {A.}~\bibnamefont {Mook}},\ }\bibfield  {title} {\bibinfo {title} {Odd-parity-wave magnons and nonrelativistic thermal $\mbox{Edelstein}$ effect},\ }\href {https://doi.org/10.48550/arXiv.2603.05415} {\bibfield  {journal} {\bibinfo  {journal} {arXiv:2603.05415}\ } (\bibinfo {year} {2026})}\BibitemShut {NoStop}%
\bibitem [{\citenamefont {Luo}\ \emph {et~al.}(2025)\citenamefont {Luo}, \citenamefont {Hu}, \citenamefont {Hu},\ and\ \citenamefont {Law}}]{LuoOdd}%
  \BibitemOpen
  \bibfield  {author} {\bibinfo {author} {\bibfnamefont {X.-J.}\ \bibnamefont {Luo}}, \bibinfo {author} {\bibfnamefont {J.-X.}\ \bibnamefont {Hu}}, \bibinfo {author} {\bibfnamefont {M.-L.}\ \bibnamefont {Hu}},\ and\ \bibinfo {author} {\bibfnamefont {K.~T.}\ \bibnamefont {Law}},\ }\bibfield  {title} {\bibinfo {title} {Spin group symmetry criteria for odd-parity magnets},\ }\href {https://arxiv.org/abs/2510.05512} {\bibfield  {journal} {\bibinfo  {journal} {arXiv:2510.05512}\ } (\bibinfo {year} {2025})}\BibitemShut {NoStop}%
\bibitem [{\citenamefont {Zeng}\ \emph {et~al.}(2026)\citenamefont {Zeng}, \citenamefont {Qin}, \citenamefont {Qin}, \citenamefont {Feng}, \citenamefont {Wu}, \citenamefont {Xu},\ and\ \citenamefont {Wang}}]{7kmk-yl2t}%
  \BibitemOpen
  \bibfield  {author} {\bibinfo {author} {\bibfnamefont {M.}~\bibnamefont {Zeng}}, \bibinfo {author} {\bibfnamefont {Z.}~\bibnamefont {Qin}}, \bibinfo {author} {\bibfnamefont {L.}~\bibnamefont {Qin}}, \bibinfo {author} {\bibfnamefont {S.}~\bibnamefont {Feng}}, \bibinfo {author} {\bibfnamefont {L.}~\bibnamefont {Wu}}, \bibinfo {author} {\bibfnamefont {D.-H.}\ \bibnamefont {Xu}},\ and\ \bibinfo {author} {\bibfnamefont {R.}~\bibnamefont {Wang}},\ }\bibfield  {title} {\bibinfo {title} {Odd-parity altermagnetism: A spin group study},\ }\href {https://doi.org/10.1103/7kmk-yl2t} {\bibfield  {journal} {\bibinfo  {journal} {Phys. Rev. B}\ }\textbf {\bibinfo {volume} {113}},\ \bibinfo {pages} {L220412} (\bibinfo {year} {2026})}\BibitemShut {NoStop}%
\bibitem [{\citenamefont {Lin}\ and\ \citenamefont {Vila}(2026)}]{c8pd-2fs4}%
  \BibitemOpen
  \bibfield  {author} {\bibinfo {author} {\bibfnamefont {Y.-P.}\ \bibnamefont {Lin}}\ and\ \bibinfo {author} {\bibfnamefont {M.}~\bibnamefont {Vila}},\ }\bibfield  {title} {\bibinfo {title} {Odd-parity altermagnetism through sublattice currents: From $\mbox{Haldane-Hubbard}$ model to general bipartite lattices},\ }\href {https://doi.org/10.1103/c8pd-2fs4} {\bibfield  {journal} {\bibinfo  {journal} {Phys. Rev. Lett.}\ }\textbf {\bibinfo {volume} {137}},\ \bibinfo {pages} {066702} (\bibinfo {year} {2026})}\BibitemShut {NoStop}%
\bibitem [{\citenamefont {Pirro}\ \emph {et~al.}(2021)\citenamefont {Pirro}, \citenamefont {Vasyuchka}, \citenamefont {Serga},\ and\ \citenamefont {Hillebrands}}]{pirro2021advances}%
  \BibitemOpen
  \bibfield  {author} {\bibinfo {author} {\bibfnamefont {P.}~\bibnamefont {Pirro}}, \bibinfo {author} {\bibfnamefont {V.~I.}\ \bibnamefont {Vasyuchka}}, \bibinfo {author} {\bibfnamefont {A.~A.}\ \bibnamefont {Serga}},\ and\ \bibinfo {author} {\bibfnamefont {B.}~\bibnamefont {Hillebrands}},\ }\bibfield  {title} {\bibinfo {title} {Advances in coherent magnonics},\ }\href {https://www.nature.com/articles/s41578-021-00332-w} {\bibfield  {journal} {\bibinfo  {journal} {Nat. Rev. Mater.}\ }\textbf {\bibinfo {volume} {6}},\ \bibinfo {pages} {1114} (\bibinfo {year} {2021})}\BibitemShut {NoStop}%
\bibitem [{\citenamefont {Hortensius}\ \emph {et~al.}(2021)\citenamefont {Hortensius}, \citenamefont {Afanasiev}, \citenamefont {Matthiesen}, \citenamefont {Leenders}, \citenamefont {Citro}, \citenamefont {Kimel}, \citenamefont {Mikhaylovskiy}, \citenamefont {Ivanov},\ and\ \citenamefont {Caviglia}}]{hortensius2021coherent}%
  \BibitemOpen
  \bibfield  {author} {\bibinfo {author} {\bibfnamefont {J.~R.}\ \bibnamefont {Hortensius}}, \bibinfo {author} {\bibfnamefont {D.}~\bibnamefont {Afanasiev}}, \bibinfo {author} {\bibfnamefont {M.}~\bibnamefont {Matthiesen}}, \bibinfo {author} {\bibfnamefont {R.}~\bibnamefont {Leenders}}, \bibinfo {author} {\bibfnamefont {R.}~\bibnamefont {Citro}}, \bibinfo {author} {\bibfnamefont {A.~V.}\ \bibnamefont {Kimel}}, \bibinfo {author} {\bibfnamefont {R.~V.}\ \bibnamefont {Mikhaylovskiy}}, \bibinfo {author} {\bibfnamefont {B.~A.}\ \bibnamefont {Ivanov}},\ and\ \bibinfo {author} {\bibfnamefont {A.~D.}\ \bibnamefont {Caviglia}},\ }\bibfield  {title} {\bibinfo {title} {Coherent spin-wave transport in an antiferromagnet},\ }\href {https://www.nature.com/articles/s41567-021-01290-4} {\bibfield  {journal} {\bibinfo  {journal} {Nat. Phys.}\ }\textbf {\bibinfo {volume} {17}},\ \bibinfo {pages} {1001} (\bibinfo {year} {2021})}\BibitemShut {NoStop}%
\bibitem [{\citenamefont {Chen}\ \emph {et~al.}(2025{\natexlab{b}})\citenamefont {Chen}, \citenamefont {Jin}, \citenamefont {Yuan}, \citenamefont {Wang}, \citenamefont {Jia}, \citenamefont {Wei}, \citenamefont {Sheng}, \citenamefont {Wang}, \citenamefont {Zhang}, \citenamefont {Liu}, \citenamefont {Yu}, \citenamefont {Ansermet}, \citenamefont {Yan},\ and\ \citenamefont {Yu}}]{chen2025observation}%
  \BibitemOpen
  \bibfield  {author} {\bibinfo {author} {\bibfnamefont {J.}~\bibnamefont {Chen}}, \bibinfo {author} {\bibfnamefont {Z.}~\bibnamefont {Jin}}, \bibinfo {author} {\bibfnamefont {R.}~\bibnamefont {Yuan}}, \bibinfo {author} {\bibfnamefont {H.}~\bibnamefont {Wang}}, \bibinfo {author} {\bibfnamefont {H.}~\bibnamefont {Jia}}, \bibinfo {author} {\bibfnamefont {W.}~\bibnamefont {Wei}}, \bibinfo {author} {\bibfnamefont {L.}~\bibnamefont {Sheng}}, \bibinfo {author} {\bibfnamefont {J.}~\bibnamefont {Wang}}, \bibinfo {author} {\bibfnamefont {Y.}~\bibnamefont {Zhang}}, \bibinfo {author} {\bibfnamefont {S.}~\bibnamefont {Liu}}, \bibinfo {author} {\bibfnamefont {D.}~\bibnamefont {Yu}}, \bibinfo {author} {\bibfnamefont {J.-P.}\ \bibnamefont {Ansermet}}, \bibinfo {author} {\bibfnamefont {P.}~\bibnamefont {Yan}},\ and\ \bibinfo {author} {\bibfnamefont {H.}~\bibnamefont {Yu}},\ }\bibfield  {title} {\bibinfo {title} {Observation of coherent gapless magnons in an antiferromagnet},\ }\href
  {https://journals.aps.org/prl/abstract/10.1103/PhysRevLett.134.056701} {\bibfield  {journal} {\bibinfo  {journal} {Phys. Rev. Lett.}\ }\textbf {\bibinfo {volume} {134}},\ \bibinfo {pages} {056701} (\bibinfo {year} {2025}{\natexlab{b}})}\BibitemShut {NoStop}%
\bibitem [{\citenamefont {Cheng}\ \emph {et~al.}(2016)\citenamefont {Cheng}, \citenamefont {Okamoto},\ and\ \citenamefont {Xiao}}]{cheng2016spin}%
  \BibitemOpen
  \bibfield  {author} {\bibinfo {author} {\bibfnamefont {R.}~\bibnamefont {Cheng}}, \bibinfo {author} {\bibfnamefont {S.}~\bibnamefont {Okamoto}},\ and\ \bibinfo {author} {\bibfnamefont {D.}~\bibnamefont {Xiao}},\ }\bibfield  {title} {\bibinfo {title} {Spin $\mbox{Nernst}$ effect of magnons in collinear antiferromagnets},\ }\href {https://journals.aps.org/prl/abstract/10.1103/PhysRevLett.117.217202} {\bibfield  {journal} {\bibinfo  {journal} {Phys. Rev. Lett.}\ }\textbf {\bibinfo {volume} {117}},\ \bibinfo {pages} {217202} (\bibinfo {year} {2016})}\BibitemShut {NoStop}%
\bibitem [{\citenamefont {Zyuzin}\ and\ \citenamefont {Kovalev}(2016)}]{zyuzin2016magnon}%
  \BibitemOpen
  \bibfield  {author} {\bibinfo {author} {\bibfnamefont {V.~A.}\ \bibnamefont {Zyuzin}}\ and\ \bibinfo {author} {\bibfnamefont {A.~A.}\ \bibnamefont {Kovalev}},\ }\bibfield  {title} {\bibinfo {title} {Magnon spin $\mbox{N}$ernst effect in antiferromagnets},\ }\href {https://journals.aps.org/prl/abstract/10.1103/PhysRevLett.117.217203} {\bibfield  {journal} {\bibinfo  {journal} {Phys. Rev. Lett.}\ }\textbf {\bibinfo {volume} {117}},\ \bibinfo {pages} {217203} (\bibinfo {year} {2016})}\BibitemShut {NoStop}%
\bibitem [{\citenamefont {Zyuzin}(2026)}]{9yz2-mqd9}%
  \BibitemOpen
  \bibfield  {author} {\bibinfo {author} {\bibfnamefont {V.~A.}\ \bibnamefont {Zyuzin}},\ }\bibfield  {title} {\bibinfo {title} {Magnon thermal {Hall} effect in collinear antiferromagnets},\ }\href {https://doi.org/10.1103/9yz2-mqd9} {\bibfield  {journal} {\bibinfo  {journal} {Phys. Rev. B}\ }\textbf {\bibinfo {volume} {113}},\ \bibinfo {pages} {094424} (\bibinfo {year} {2026})}\BibitemShut {NoStop}%
\bibitem [{\citenamefont {Du}\ \emph {et~al.}(2025)\citenamefont {Du}, \citenamefont {Zhang}, \citenamefont {Ni}, \citenamefont {Jiang},\ and\ \citenamefont {Bellaiche}}]{du2025nonreciprocal}%
  \BibitemOpen
  \bibfield  {author} {\bibinfo {author} {\bibfnamefont {Q.}~\bibnamefont {Du}}, \bibinfo {author} {\bibfnamefont {Z.}~\bibnamefont {Zhang}}, \bibinfo {author} {\bibfnamefont {J.}~\bibnamefont {Ni}}, \bibinfo {author} {\bibfnamefont {Z.}~\bibnamefont {Jiang}},\ and\ \bibinfo {author} {\bibfnamefont {L.}~\bibnamefont {Bellaiche}},\ }\bibfield  {title} {\bibinfo {title} {Nonreciprocal magnons in layered antiferromagnets $\mbox{VPX}_{3}$ ($\mbox{X}$\,=\,$\mbox{S}$,\,$\mbox{Se}$,\,$\mbox{Te}$)},\ }\href {https://journals.aps.org/prb/abstract/10.1103/8fx7-4xgm} {\bibfield  {journal} {\bibinfo  {journal} {Phys. Rev. B}\ }\textbf {\bibinfo {volume} {112}},\ \bibinfo {pages} {L100408} (\bibinfo {year} {2025})}\BibitemShut {NoStop}%
\bibitem [{\citenamefont {Ni}\ \emph {et~al.}(2025)\citenamefont {Ni}, \citenamefont {Jin}, \citenamefont {Du},\ and\ \citenamefont {Chang}}]{ni2025magnon}%
  \BibitemOpen
  \bibfield  {author} {\bibinfo {author} {\bibfnamefont {J.}~\bibnamefont {Ni}}, \bibinfo {author} {\bibfnamefont {Y.}~\bibnamefont {Jin}}, \bibinfo {author} {\bibfnamefont {Q.}~\bibnamefont {Du}},\ and\ \bibinfo {author} {\bibfnamefont {G.}~\bibnamefont {Chang}},\ }\bibfield  {title} {\bibinfo {title} {Magnon nonlinear $\mbox{Hall}$ effect in two-dimensional antiferromagnetic insulators},\ }\href {https://journals.aps.org/prb/abstract/10.1103/72sf-6km4} {\bibfield  {journal} {\bibinfo  {journal} {Phys. Rev. B}\ }\textbf {\bibinfo {volume} {112}},\ \bibinfo {pages} {054424} (\bibinfo {year} {2025})}\BibitemShut {NoStop}%
\bibitem [{\citenamefont {Bai}\ \emph {et~al.}(2025)\citenamefont {Bai}, \citenamefont {Zhang}, \citenamefont {Feng},\ and\ \citenamefont {Yao}}]{rn1l-d6cq}%
  \BibitemOpen
  \bibfield  {author} {\bibinfo {author} {\bibfnamefont {L.}~\bibnamefont {Bai}}, \bibinfo {author} {\bibfnamefont {R.-W.}\ \bibnamefont {Zhang}}, \bibinfo {author} {\bibfnamefont {W.}~\bibnamefont {Feng}},\ and\ \bibinfo {author} {\bibfnamefont {Y.}~\bibnamefont {Yao}},\ }\bibfield  {title} {\bibinfo {title} {Anomalous $\mbox{Hall}$ effect in type {IV} $\mbox{2D}$ collinear magnets},\ }\href {https://doi.org/10.1103/rn1l-d6cq} {\bibfield  {journal} {\bibinfo  {journal} {Phys. Rev. Lett.}\ }\textbf {\bibinfo {volume} {135}},\ \bibinfo {pages} {036702} (\bibinfo {year} {2025})}\BibitemShut {NoStop}%
\bibitem [{\citenamefont {Liu}\ \emph {et~al.}(2025{\natexlab{a}})\citenamefont {Liu}, \citenamefont {Guo}, \citenamefont {Li},\ and\ \citenamefont {Liu}}]{liu20252D}%
  \BibitemOpen
  \bibfield  {author} {\bibinfo {author} {\bibfnamefont {Y.}~\bibnamefont {Liu}}, \bibinfo {author} {\bibfnamefont {S.-D.}\ \bibnamefont {Guo}}, \bibinfo {author} {\bibfnamefont {Y.}~\bibnamefont {Li}},\ and\ \bibinfo {author} {\bibfnamefont {C.-C.}\ \bibnamefont {Liu}},\ }\bibfield  {title} {\bibinfo {title} {Two-dimensional fully compensated ferrimagnetism},\ }\href {https://doi.org/10.1103/PhysRevLett.134.116703} {\bibfield  {journal} {\bibinfo  {journal} {Phys. Rev. Lett.}\ }\textbf {\bibinfo {volume} {134}},\ \bibinfo {pages} {116703} (\bibinfo {year} {2025}{\natexlab{a}})}\BibitemShut {NoStop}%
\bibitem [{\citenamefont {Holstein}\ and\ \citenamefont {Primakoff}(1940)}]{holstein1940field}%
  \BibitemOpen
  \bibfield  {author} {\bibinfo {author} {\bibfnamefont {T.}~\bibnamefont {Holstein}}\ and\ \bibinfo {author} {\bibfnamefont {H.}~\bibnamefont {Primakoff}},\ }\bibfield  {title} {\bibinfo {title} {Field dependence of the intrinsic domain magnetization of a ferromagnet},\ }\href {https://journals.aps.org/pr/abstract/10.1103/PhysRev.58.1098} {\bibfield  {journal} {\bibinfo  {journal} {Phys. Rev.}\ }\textbf {\bibinfo {volume} {58}},\ \bibinfo {pages} {1098} (\bibinfo {year} {1940})}\BibitemShut {NoStop}%
\bibitem [{Sup()}]{Supplemental_Materials}%
  \BibitemOpen
  \href@noop {} {\bibinfo  {journal} {See Supplemental Material at ... for (i) details of the calculations; and (ii) the Hamiltonians, eigenvalues, eigenvectors, and Berry curvatures for $p$-wave, $f$-wave, and nodeless chiral magnons.}\ }\BibitemShut {NoStop}%
\bibitem [{\citenamefont {Bogoljubov}\ \emph {et~al.}(1958)\citenamefont {Bogoljubov}, \citenamefont {Tolmachov},\ and\ \citenamefont {{\v{S}}irkov}}]{bogoljubov1958new}%
  \BibitemOpen
\bibfield  {journal} {  }\bibfield  {author} {\bibinfo {author} {\bibfnamefont {N.~N.}\ \bibnamefont {Bogoljubov}}, \bibinfo {author} {\bibfnamefont {V.~V.}\ \bibnamefont {Tolmachov}},\ and\ \bibinfo {author} {\bibfnamefont {D.~V.}\ \bibnamefont {{\v{S}}irkov}},\ }\bibfield  {title} {\bibinfo {title} {A new method in the theory of superconductivity},\ }\href {https://onlinelibrary.wiley.com/doi/abs/10.1002/prop.19580061102} {\bibfield  {journal} {\bibinfo  {journal} {Fortschr. Phys.}\ }\textbf {\bibinfo {volume} {6}},\ \bibinfo {pages} {605} (\bibinfo {year} {1958})}\BibitemShut {NoStop}%
\bibitem [{\citenamefont {Valatin}(1958)}]{valatin1958comments}%
  \BibitemOpen
  \bibfield  {author} {\bibinfo {author} {\bibfnamefont {J.~G.}\ \bibnamefont {Valatin}},\ }\bibfield  {title} {\bibinfo {title} {Comments on the theory of superconductivity},\ }\href {https://link.springer.com/article/10.1007/BF02745589} {\bibfield  {journal} {\bibinfo  {journal} {Nuovo Cim.}\ }\textbf {\bibinfo {volume} {7}},\ \bibinfo {pages} {843} (\bibinfo {year} {1958})}\BibitemShut {NoStop}%
\bibitem [{\citenamefont {Matsumoto}\ and\ \citenamefont {Hayami}(2020)}]{matsumoto2020nonreciprocal}%
  \BibitemOpen
  \bibfield  {author} {\bibinfo {author} {\bibfnamefont {T.}~\bibnamefont {Matsumoto}}\ and\ \bibinfo {author} {\bibfnamefont {S.}~\bibnamefont {Hayami}},\ }\bibfield  {title} {\bibinfo {title} {Nonreciprocal magnons due to symmetric anisotropic exchange interaction in honeycomb antiferromagnets},\ }\href {https://journals.aps.org/prb/abstract/10.1103/PhysRevB.101.224419} {\bibfield  {journal} {\bibinfo  {journal} {Phys. Rev. B}\ }\textbf {\bibinfo {volume} {101}},\ \bibinfo {pages} {224419} (\bibinfo {year} {2020})}\BibitemShut {NoStop}%
\bibitem [{\citenamefont {Sato}\ and\ \citenamefont {Matan}(2019)}]{sato2019nonreciprocal}%
  \BibitemOpen
  \bibfield  {author} {\bibinfo {author} {\bibfnamefont {T.~J.}\ \bibnamefont {Sato}}\ and\ \bibinfo {author} {\bibfnamefont {K.}~\bibnamefont {Matan}},\ }\bibfield  {title} {\bibinfo {title} {Nonreciprocal magnons in noncentrosymmetric magnets},\ }\href {https://journals.jps.jp/doi/10.7566/JPSJ.88.081007} {\bibfield  {journal} {\bibinfo  {journal} {J. Phys. Soc. Jpn.}\ }\textbf {\bibinfo {volume} {88}},\ \bibinfo {pages} {081007} (\bibinfo {year} {2019})}\BibitemShut {NoStop}%
\bibitem [{\citenamefont {Gitgeatpong}\ \emph {et~al.}(2017)\citenamefont {Gitgeatpong}, \citenamefont {Zhao}, \citenamefont {Piyawongwatthana}, \citenamefont {Qiu}, \citenamefont {Harriger}, \citenamefont {Butch}, \citenamefont {Sato},\ and\ \citenamefont {Matan}}]{gitgeatpong2017nonreciprocal}%
  \BibitemOpen
  \bibfield  {author} {\bibinfo {author} {\bibfnamefont {G.}~\bibnamefont {Gitgeatpong}}, \bibinfo {author} {\bibfnamefont {Y.}~\bibnamefont {Zhao}}, \bibinfo {author} {\bibfnamefont {P.}~\bibnamefont {Piyawongwatthana}}, \bibinfo {author} {\bibfnamefont {Y.}~\bibnamefont {Qiu}}, \bibinfo {author} {\bibfnamefont {L.~W.}\ \bibnamefont {Harriger}}, \bibinfo {author} {\bibfnamefont {N.~P.}\ \bibnamefont {Butch}}, \bibinfo {author} {\bibfnamefont {T.~J.}\ \bibnamefont {Sato}},\ and\ \bibinfo {author} {\bibfnamefont {K.}~\bibnamefont {Matan}},\ }\bibfield  {title} {\bibinfo {title} {Nonreciprocal magnons and symmetry-breaking in the noncentrosymmetric antiferromagnet},\ }\href {https://journals.aps.org/prl/abstract/10.1103/PhysRevLett.119.047201} {\bibfield  {journal} {\bibinfo  {journal} {Phys. Rev. Lett.}\ }\textbf {\bibinfo {volume} {119}},\ \bibinfo {pages} {047201} (\bibinfo {year} {2017})}\BibitemShut {NoStop}%
\bibitem [{\citenamefont {Kawano}\ \emph {et~al.}(2019)\citenamefont {Kawano}, \citenamefont {Onose},\ and\ \citenamefont {Hotta}}]{kawano2019designing}%
  \BibitemOpen
  \bibfield  {author} {\bibinfo {author} {\bibfnamefont {M.}~\bibnamefont {Kawano}}, \bibinfo {author} {\bibfnamefont {Y.}~\bibnamefont {Onose}},\ and\ \bibinfo {author} {\bibfnamefont {C.}~\bibnamefont {Hotta}},\ }\bibfield  {title} {\bibinfo {title} {Designing {Rashba--Dresselhaus} effect in magnetic insulators},\ }\href {https://doi.org/10.1038/s42005-019-0128-6} {\bibfield  {journal} {\bibinfo  {journal} {Commun. Phys.}\ }\textbf {\bibinfo {volume} {2}},\ \bibinfo {pages} {27} (\bibinfo {year} {2019})}\BibitemShut {NoStop}%
\bibitem [{\citenamefont {Okuma}(2017)}]{okuma2017magnon}%
  \BibitemOpen
  \bibfield  {author} {\bibinfo {author} {\bibfnamefont {N.}~\bibnamefont {Okuma}},\ }\bibfield  {title} {\bibinfo {title} {Magnon spin-momentum locking: Various spin vortices and {Dirac} magnons in noncollinear antiferromagnets},\ }\href {https://doi.org/10.1103/PhysRevLett.119.107205} {\bibfield  {journal} {\bibinfo  {journal} {Phys. Rev. Lett.}\ }\textbf {\bibinfo {volume} {119}},\ \bibinfo {pages} {107205} (\bibinfo {year} {2017})}\BibitemShut {NoStop}%
\bibitem [{\citenamefont {Dzyaloshinsky}(1958)}]{dzyaloshinsky_JPCS_1958_4}%
  \BibitemOpen
  \bibfield  {author} {\bibinfo {author} {\bibfnamefont {I.}~\bibnamefont {Dzyaloshinsky}},\ }\bibfield  {title} {\bibinfo {title} {A thermodynamic theory of “weak” ferromagnetism of antiferromagnetics},\ }\href {https://www.sciencedirect.com/science/article/abs/pii/0022369758900763} {\bibfield  {journal} {\bibinfo  {journal} {J. Phys. Chem. Solids}\ }\textbf {\bibinfo {volume} {4}},\ \bibinfo {pages} {241} (\bibinfo {year} {1958})}\BibitemShut {NoStop}%
\bibitem [{\citenamefont {Moriya}(1960)}]{moriya_PR_1960_120}%
  \BibitemOpen
  \bibfield  {author} {\bibinfo {author} {\bibfnamefont {T.}~\bibnamefont {Moriya}},\ }\bibfield  {title} {\bibinfo {title} {Anisotropic superexchange interaction and weak ferromagnetism},\ }\href {https://journals.aps.org/pr/abstract/10.1103/PhysRev.120.91} {\bibfield  {journal} {\bibinfo  {journal} {Phys. Rev.}\ }\textbf {\bibinfo {volume} {120}},\ \bibinfo {pages} {91} (\bibinfo {year} {1960})}\BibitemShut {NoStop}%
\bibitem [{\citenamefont {Zhao}\ \emph {et~al.}(2021)\citenamefont {Zhao}, \citenamefont {Chen}, \citenamefont {Prosandeev}, \citenamefont {Artyukhin},\ and\ \citenamefont {Bellaiche}}]{zhao2021dzyaloshinskii}%
  \BibitemOpen
  \bibfield  {author} {\bibinfo {author} {\bibfnamefont {H.~J.}\ \bibnamefont {Zhao}}, \bibinfo {author} {\bibfnamefont {P.}~\bibnamefont {Chen}}, \bibinfo {author} {\bibfnamefont {S.}~\bibnamefont {Prosandeev}}, \bibinfo {author} {\bibfnamefont {S.}~\bibnamefont {Artyukhin}},\ and\ \bibinfo {author} {\bibfnamefont {L.}~\bibnamefont {Bellaiche}},\ }\bibfield  {title} {\bibinfo {title} {$\mbox{Dzyaloshinskii--Moriya-like}$ interaction in ferroelectrics and antiferroelectrics},\ }\href {https://www.nature.com/articles/s41563-020-00821-3} {\bibfield  {journal} {\bibinfo  {journal} {Nat. Mater.}\ }\textbf {\bibinfo {volume} {20}},\ \bibinfo {pages} {341} (\bibinfo {year} {2021})}\BibitemShut {NoStop}%
\bibitem [{\citenamefont {Zhao}\ \emph {et~al.}(2014)\citenamefont {Zhao}, \citenamefont {Ren}, \citenamefont {Yang}, \citenamefont {{\'I}{\~n}iguez}, \citenamefont {Chen},\ and\ \citenamefont {Bellaiche}}]{zhao2014near}%
  \BibitemOpen
  \bibfield  {author} {\bibinfo {author} {\bibfnamefont {H.~J.}\ \bibnamefont {Zhao}}, \bibinfo {author} {\bibfnamefont {W.}~\bibnamefont {Ren}}, \bibinfo {author} {\bibfnamefont {Y.}~\bibnamefont {Yang}}, \bibinfo {author} {\bibfnamefont {J.}~\bibnamefont {{\'I}{\~n}iguez}}, \bibinfo {author} {\bibfnamefont {X.~M.}\ \bibnamefont {Chen}},\ and\ \bibinfo {author} {\bibfnamefont {L.}~\bibnamefont {Bellaiche}},\ }\bibfield  {title} {\bibinfo {title} {Near room-temperature multiferroic materials with tunable ferromagnetic and electrical properties},\ }\href {https://www.nature.com/articles/ncomms5021} {\bibfield  {journal} {\bibinfo  {journal} {Nat. Commun.}\ }\textbf {\bibinfo {volume} {5}},\ \bibinfo {pages} {4021} (\bibinfo {year} {2014})}\BibitemShut {NoStop}%
\bibitem [{\citenamefont {Xiao}\ \emph {et~al.}(2010)\citenamefont {Xiao}, \citenamefont {Chang},\ and\ \citenamefont {Niu}}]{xiao2010berry}%
  \BibitemOpen
  \bibfield  {author} {\bibinfo {author} {\bibfnamefont {D.}~\bibnamefont {Xiao}}, \bibinfo {author} {\bibfnamefont {M.-C.}\ \bibnamefont {Chang}},\ and\ \bibinfo {author} {\bibfnamefont {Q.}~\bibnamefont {Niu}},\ }\bibfield  {title} {\bibinfo {title} {Berry phase effects on electronic properties},\ }\href {https://journals.aps.org/rmp/abstract/10.1103/RevModPhys.82.1959} {\bibfield  {journal} {\bibinfo  {journal} {Rev. Mod. Phys.}\ }\textbf {\bibinfo {volume} {82}},\ \bibinfo {pages} {1959} (\bibinfo {year} {2010})}\BibitemShut {NoStop}%
\bibitem [{\citenamefont {Brinkman}\ and\ \citenamefont {Elliott}(1966)}]{brinkman1966theory}%
  \BibitemOpen
  \bibfield  {author} {\bibinfo {author} {\bibfnamefont {W.~F.}\ \bibnamefont {Brinkman}}\ and\ \bibinfo {author} {\bibfnamefont {R.~J.}\ \bibnamefont {Elliott}},\ }\bibfield  {title} {\bibinfo {title} {Theory of spin-space groups},\ }\href {https://royalsocietypublishing.org/doi/10.1098/rspa.1966.0211} {\bibfield  {journal} {\bibinfo  {journal} {Proc. R. Soc. A}\ }\textbf {\bibinfo {volume} {294}},\ \bibinfo {pages} {343} (\bibinfo {year} {1966})}\BibitemShut {NoStop}%
\bibitem [{\citenamefont {Matsumoto}\ and\ \citenamefont {Murakami}(2011)}]{matsumoto2011theoretical}%
  \BibitemOpen
  \bibfield  {author} {\bibinfo {author} {\bibfnamefont {R.}~\bibnamefont {Matsumoto}}\ and\ \bibinfo {author} {\bibfnamefont {S.}~\bibnamefont {Murakami}},\ }\bibfield  {title} {\bibinfo {title} {Theoretical prediction of a rotating magnon wave packet in ferromagnets},\ }\href {https://journals.aps.org/prl/abstract/10.1103/PhysRevLett.106.197202} {\bibfield  {journal} {\bibinfo  {journal} {Phys. Rev. Lett.}\ }\textbf {\bibinfo {volume} {106}},\ \bibinfo {pages} {197202} (\bibinfo {year} {2011})}\BibitemShut {NoStop}%
\bibitem [{\citenamefont {Wu}\ \emph {et~al.}(2025)\citenamefont {Wu}, \citenamefont {Dong}, \citenamefont {Zhu}, \citenamefont {Zheng},\ and\ \citenamefont {Zhang}}]{wu2025magnon}%
  \BibitemOpen
  \bibfield  {author} {\bibinfo {author} {\bibfnamefont {K.}~\bibnamefont {Wu}}, \bibinfo {author} {\bibfnamefont {J.}~\bibnamefont {Dong}}, \bibinfo {author} {\bibfnamefont {M.}~\bibnamefont {Zhu}}, \bibinfo {author} {\bibfnamefont {F.}~\bibnamefont {Zheng}},\ and\ \bibinfo {author} {\bibfnamefont {J.}~\bibnamefont {Zhang}},\ }\bibfield  {title} {\bibinfo {title} {Magnon splitting and magnon spin transport in altermagnets},\ }\href {https://iopscience.iop.org/article/10.1088/0256-307X/42/7/070702/meta} {\bibfield  {journal} {\bibinfo  {journal} {Chin. Phys. Lett.}\ }\textbf {\bibinfo {volume} {42}},\ \bibinfo {pages} {070702} (\bibinfo {year} {2025})}\BibitemShut {NoStop}%
\bibitem [{\citenamefont {Li}\ \emph {et~al.}(2020)\citenamefont {Li}, \citenamefont {Mook}, \citenamefont {Raeliarijaona},\ and\ \citenamefont {Kovalev}}]{libo2020}%
  \BibitemOpen
  \bibfield  {author} {\bibinfo {author} {\bibfnamefont {B.}~\bibnamefont {Li}}, \bibinfo {author} {\bibfnamefont {A.}~\bibnamefont {Mook}}, \bibinfo {author} {\bibfnamefont {A.}~\bibnamefont {Raeliarijaona}},\ and\ \bibinfo {author} {\bibfnamefont {A.~A.}\ \bibnamefont {Kovalev}},\ }\bibfield  {title} {\bibinfo {title} {Magnonic analog of the $\mbox{Edelstein}$ effect in antiferromagnetic insulators},\ }\href {https://doi.org/10.1103/PhysRevB.101.024427} {\bibfield  {journal} {\bibinfo  {journal} {Phys. Rev. B}\ }\textbf {\bibinfo {volume} {101}},\ \bibinfo {pages} {024427} (\bibinfo {year} {2020})}\BibitemShut {NoStop}%
\bibitem [{\citenamefont {Grzybowski}\ \emph {et~al.}(2024)\citenamefont {Grzybowski}, \citenamefont {Autieri}, \citenamefont {Domagala}, \citenamefont {Krasucki}, \citenamefont {Kaleta}, \citenamefont {Kret}, \citenamefont {Gas}, \citenamefont {Sawicki}, \citenamefont {Bo{\.z}ek}, \citenamefont {Suffczy{\'n}ski},\ and\ \citenamefont {Pacuski}}]{grzybowski2024wurtzite}%
  \BibitemOpen
  \bibfield  {author} {\bibinfo {author} {\bibfnamefont {M.~J.}\ \bibnamefont {Grzybowski}}, \bibinfo {author} {\bibfnamefont {C.}~\bibnamefont {Autieri}}, \bibinfo {author} {\bibfnamefont {J.}~\bibnamefont {Domagala}}, \bibinfo {author} {\bibfnamefont {C.}~\bibnamefont {Krasucki}}, \bibinfo {author} {\bibfnamefont {A.}~\bibnamefont {Kaleta}}, \bibinfo {author} {\bibfnamefont {S.}~\bibnamefont {Kret}}, \bibinfo {author} {\bibfnamefont {K.}~\bibnamefont {Gas}}, \bibinfo {author} {\bibfnamefont {M.}~\bibnamefont {Sawicki}}, \bibinfo {author} {\bibfnamefont {R.}~\bibnamefont {Bo{\.z}ek}}, \bibinfo {author} {\bibfnamefont {J.}~\bibnamefont {Suffczy{\'n}ski}},\ and\ \bibinfo {author} {\bibfnamefont {W.}~\bibnamefont {Pacuski}},\ }\bibfield  {title} {\bibinfo {title} {Wurtzite $vs.$ rock-salt $\mbox{MnSe}$ epitaxy: electronic and altermagnetic properties},\ }\href {https://pubs.rsc.org/en/content/articlehtml/2024/nr/d3nr04798a} {\bibfield  {journal} {\bibinfo  {journal} {Nanoscale}\ }\textbf {\bibinfo {volume}
  {16}},\ \bibinfo {pages} {6259} (\bibinfo {year} {2024})}\BibitemShut {NoStop}%
\bibitem [{\citenamefont {Bezzerga}\ \emph {et~al.}(2025)\citenamefont {Bezzerga}, \citenamefont {Khan},\ and\ \citenamefont {Hong}}]{bezzerga2025high}%
  \BibitemOpen
  \bibfield  {author} {\bibinfo {author} {\bibfnamefont {D.}~\bibnamefont {Bezzerga}}, \bibinfo {author} {\bibfnamefont {I.}~\bibnamefont {Khan}},\ and\ \bibinfo {author} {\bibfnamefont {J.}~\bibnamefont {Hong}},\ }\bibfield  {title} {\bibinfo {title} {High performance room temperature multiferroic properties of $w$-$\mbox{MnSe}$ altermagnet},\ }\href {https://advanced.onlinelibrary.wiley.com/doi/full/10.1002/adfm.202505813} {\bibfield  {journal} {\bibinfo  {journal} {Adv. Funct. Mater.}\ }\textbf {\bibinfo {volume} {35}},\ \bibinfo {pages} {2505813} (\bibinfo {year} {2025})}\BibitemShut {NoStop}%
\bibitem [{\citenamefont {Van~Aken}\ \emph {et~al.}(2004)\citenamefont {Van~Aken}, \citenamefont {Palstra}, \citenamefont {Filippetti},\ and\ \citenamefont {Spaldin}}]{van2004origin}%
  \BibitemOpen
  \bibfield  {author} {\bibinfo {author} {\bibfnamefont {B.~B.}\ \bibnamefont {Van~Aken}}, \bibinfo {author} {\bibfnamefont {T.~T.~M.}\ \bibnamefont {Palstra}}, \bibinfo {author} {\bibfnamefont {A.}~\bibnamefont {Filippetti}},\ and\ \bibinfo {author} {\bibfnamefont {N.~A.}\ \bibnamefont {Spaldin}},\ }\bibfield  {title} {\bibinfo {title} {The origin of ferroelectricity in magnetoelectric $\textrm{YMnO}_{3}$},\ }\href {https://doi.org/10.1038/nmat1080} {\bibfield  {journal} {\bibinfo  {journal} {Nat. Mater.}\ }\textbf {\bibinfo {volume} {3}},\ \bibinfo {pages} {164} (\bibinfo {year} {2004})}\BibitemShut {NoStop}%
\bibitem [{\citenamefont {Zhou}\ \emph {et~al.}(2025{\natexlab{b}})\citenamefont {Zhou}, \citenamefont {Zhang}, \citenamefont {Ji}, \citenamefont {Xiang}, \citenamefont {Dong}, \citenamefont {Rondinelli},\ and\ \citenamefont {Lu}}]{xpsy-cjn6}%
  \BibitemOpen
  \bibfield  {author} {\bibinfo {author} {\bibfnamefont {Y.}~\bibnamefont {Zhou}}, \bibinfo {author} {\bibfnamefont {H.-M.}\ \bibnamefont {Zhang}}, \bibinfo {author} {\bibfnamefont {C.-A.}\ \bibnamefont {Ji}}, \bibinfo {author} {\bibfnamefont {H.}~\bibnamefont {Xiang}}, \bibinfo {author} {\bibfnamefont {S.}~\bibnamefont {Dong}}, \bibinfo {author} {\bibfnamefont {J.~M.}\ \bibnamefont {Rondinelli}},\ and\ \bibinfo {author} {\bibfnamefont {X.-Z.}\ \bibnamefont {Lu}},\ }\bibfield  {title} {\bibinfo {title} {Piezomagnetism-driven magnetoelectric coupling in altermagnetic multiferroic {${\mbox{K}}_{3}{\mbox{Cr}}_{2}{\mbox{F}}_{7}$}},\ }\href {https://doi.org/10.1103/xpsy-cjn6} {\bibfield  {journal} {\bibinfo  {journal} {Phys. Rev. B}\ }\textbf {\bibinfo {volume} {112}},\ \bibinfo {pages} {094412} (\bibinfo {year} {2025}{\natexlab{b}})}\BibitemShut {NoStop}%
\bibitem [{\citenamefont {Xu}\ \emph {et~al.}(2017)\citenamefont {Xu}, \citenamefont {Lu},\ and\ \citenamefont {Xiang}}]{xu2017designing}%
  \BibitemOpen
  \bibfield  {author} {\bibinfo {author} {\bibfnamefont {K.}~\bibnamefont {Xu}}, \bibinfo {author} {\bibfnamefont {X.-Z.}\ \bibnamefont {Lu}},\ and\ \bibinfo {author} {\bibfnamefont {H.}~\bibnamefont {Xiang}},\ }\bibfield  {title} {\bibinfo {title} {Designing new ferroelectrics with a general strategy},\ }\href {https://www.nature.com/articles/s41535-016-0001-8} {\bibfield  {journal} {\bibinfo  {journal} {npj Quant. Mater.}\ }\textbf {\bibinfo {volume} {2}},\ \bibinfo {pages} {1} (\bibinfo {year} {2017})}\BibitemShut {NoStop}%
\bibitem [{\citenamefont {Zhu}\ \emph {et~al.}(2025)\citenamefont {Zhu}, \citenamefont {Duan}, \citenamefont {Zhang}, \citenamefont {Hao}, \citenamefont {Žutić},\ and\ \citenamefont {Zhou}}]{zhu2025two}%
  \BibitemOpen
  \bibfield  {author} {\bibinfo {author} {\bibfnamefont {Z.}~\bibnamefont {Zhu}}, \bibinfo {author} {\bibfnamefont {X.}~\bibnamefont {Duan}}, \bibinfo {author} {\bibfnamefont {J.}~\bibnamefont {Zhang}}, \bibinfo {author} {\bibfnamefont {B.}~\bibnamefont {Hao}}, \bibinfo {author} {\bibfnamefont {I.}~\bibnamefont {Žutić}},\ and\ \bibinfo {author} {\bibfnamefont {T.}~\bibnamefont {Zhou}},\ }\bibfield  {title} {\bibinfo {title} {Two-dimensional ferroelectric altermagnets: From model to material realization},\ }\href {https://doi.org/10.1021/acs.nanolett.5c02121} {\bibfield  {journal} {\bibinfo  {journal} {Nano Lett.}\ }\textbf {\bibinfo {volume} {25}},\ \bibinfo {pages} {9456} (\bibinfo {year} {2025})}\BibitemShut {NoStop}%
\bibitem [{\citenamefont {Tan}\ \emph {et~al.}(2019)\citenamefont {Tan}, \citenamefont {Li}, \citenamefont {Liu}, \citenamefont {Liu}, \citenamefont {Li},\ and\ \citenamefont {Duan}}]{Tan2019}%
  \BibitemOpen
  \bibfield  {author} {\bibinfo {author} {\bibfnamefont {H.}~\bibnamefont {Tan}}, \bibinfo {author} {\bibfnamefont {M.}~\bibnamefont {Li}}, \bibinfo {author} {\bibfnamefont {H.}~\bibnamefont {Liu}}, \bibinfo {author} {\bibfnamefont {Z.}~\bibnamefont {Liu}}, \bibinfo {author} {\bibfnamefont {Y.}~\bibnamefont {Li}},\ and\ \bibinfo {author} {\bibfnamefont {W.}~\bibnamefont {Duan}},\ }\bibfield  {title} {\bibinfo {title} {Two-dimensional ferromagnetic-ferroelectric multiferroics in violation of the ${d}^{0}$ rule},\ }\href {https://doi.org/10.1103/PhysRevB.99.195434} {\bibfield  {journal} {\bibinfo  {journal} {Phys. Rev. B}\ }\textbf {\bibinfo {volume} {99}},\ \bibinfo {pages} {195434} (\bibinfo {year} {2019})}\BibitemShut {NoStop}%
\bibitem [{\citenamefont {Kravchuk}\ \emph {et~al.}(2026)\citenamefont {Kravchuk}, \citenamefont {Yershov}, \citenamefont {Pradenas}, \citenamefont {Neumann}, \citenamefont {Jaeschke-Ubiergo}, \citenamefont {Zarzuela}, \citenamefont {Sinova}, \citenamefont {van~den Brink},\ and\ \citenamefont {Mook}}]{kravchuk2026nonlinear}%
  \BibitemOpen
  \bibfield  {author} {\bibinfo {author} {\bibfnamefont {V.~P.}\ \bibnamefont {Kravchuk}}, \bibinfo {author} {\bibfnamefont {K.~V.}\ \bibnamefont {Yershov}}, \bibinfo {author} {\bibfnamefont {B.}~\bibnamefont {Pradenas}}, \bibinfo {author} {\bibfnamefont {R.~R.}\ \bibnamefont {Neumann}}, \bibinfo {author} {\bibfnamefont {R.}~\bibnamefont {Jaeschke-Ubiergo}}, \bibinfo {author} {\bibfnamefont {R.}~\bibnamefont {Zarzuela}}, \bibinfo {author} {\bibfnamefont {J.}~\bibnamefont {Sinova}}, \bibinfo {author} {\bibfnamefont {J.}~\bibnamefont {van~den Brink}},\ and\ \bibinfo {author} {\bibfnamefont {A.}~\bibnamefont {Mook}},\ }\bibfield  {title} {\bibinfo {title} {Nonlinear magnon magnetic moment transport in triangular-lattice f-wave antialtermagnets},\ }\href {https://arxiv.org/abs/2605.22614} {\bibfield  {journal} {\bibinfo  {journal} {arXiv:2605.22614}\ } (\bibinfo {year} {2026})}\BibitemShut {NoStop}%
\bibitem [{\citenamefont {Gu}\ \emph {et~al.}(2025)\citenamefont {Gu}, \citenamefont {Liu}, \citenamefont {Zhu}, \citenamefont {Yananose}, \citenamefont {Chen}, \citenamefont {Hu}, \citenamefont {Stroppa},\ and\ \citenamefont {Liu}}]{Gumingqiang025}%
  \BibitemOpen
  \bibfield  {author} {\bibinfo {author} {\bibfnamefont {M.}~\bibnamefont {Gu}}, \bibinfo {author} {\bibfnamefont {Y.}~\bibnamefont {Liu}}, \bibinfo {author} {\bibfnamefont {H.}~\bibnamefont {Zhu}}, \bibinfo {author} {\bibfnamefont {K.}~\bibnamefont {Yananose}}, \bibinfo {author} {\bibfnamefont {X.}~\bibnamefont {Chen}}, \bibinfo {author} {\bibfnamefont {Y.}~\bibnamefont {Hu}}, \bibinfo {author} {\bibfnamefont {A.}~\bibnamefont {Stroppa}},\ and\ \bibinfo {author} {\bibfnamefont {Q.}~\bibnamefont {Liu}},\ }\bibfield  {title} {\bibinfo {title} {Ferroelectric switchable altermagnetism},\ }\href {https://doi.org/10.1103/PhysRevLett.134.106802} {\bibfield  {journal} {\bibinfo  {journal} {Phys. Rev. Lett.}\ }\textbf {\bibinfo {volume} {134}},\ \bibinfo {pages} {106802} (\bibinfo {year} {2025})}\BibitemShut {NoStop}%
\bibitem [{\citenamefont {Zhao}\ \emph {et~al.}(2026)\citenamefont {Zhao}, \citenamefont {Zhou}, \citenamefont {Guo}, \citenamefont {Zhu}, \citenamefont {Chen}, \citenamefont {Li}, \citenamefont {Cheng}, \citenamefont {Wang},\ and\ \citenamefont {Wang}}]{zhao2026multiferroic}%
  \BibitemOpen
  \bibfield  {author} {\bibinfo {author} {\bibfnamefont {W.}~\bibnamefont {Zhao}}, \bibinfo {author} {\bibfnamefont {X.}~\bibnamefont {Zhou}}, \bibinfo {author} {\bibfnamefont {Z.}~\bibnamefont {Guo}}, \bibinfo {author} {\bibfnamefont {T.}~\bibnamefont {Zhu}}, \bibinfo {author} {\bibfnamefont {J.}~\bibnamefont {Chen}}, \bibinfo {author} {\bibfnamefont {H.}~\bibnamefont {Li}}, \bibinfo {author} {\bibfnamefont {Z.}~\bibnamefont {Cheng}}, \bibinfo {author} {\bibfnamefont {X.}~\bibnamefont {Wang}},\ and\ \bibinfo {author} {\bibfnamefont {W.}~\bibnamefont {Wang}},\ }\bibfield  {title} {\bibinfo {title} {Multiferroic phase transition between multiple types of collinear compensated magnets},\ }\href {https://doi.org/10.1038/s41467-026-72339-2} {\bibfield  {journal} {\bibinfo  {journal} {Nat. Commun.}\ }\textbf {\bibinfo {volume} {17}},\ \bibinfo {pages} {5663} (\bibinfo {year} {2026})}\BibitemShut {NoStop}%
\bibitem [{\citenamefont {Sun}\ \emph {et~al.}(2026)\citenamefont {Sun}, \citenamefont {Yang}, \citenamefont {Wang}, \citenamefont {Huang},\ and\ \citenamefont {Cheng}}]{sun2026altermagnetic}%
  \BibitemOpen
  \bibfield  {author} {\bibinfo {author} {\bibfnamefont {W.}~\bibnamefont {Sun}}, \bibinfo {author} {\bibfnamefont {C.}~\bibnamefont {Yang}}, \bibinfo {author} {\bibfnamefont {X.}~\bibnamefont {Wang}}, \bibinfo {author} {\bibfnamefont {S.}~\bibnamefont {Huang}},\ and\ \bibinfo {author} {\bibfnamefont {Z.}~\bibnamefont {Cheng}},\ }\bibfield  {title} {\bibinfo {title} {Altermagnetic multiferroics with symmetry-locked magnetoelectric coupling},\ }\href {https://doi.org/10.1038/s41563-026-02518-5} {\bibfield  {journal} {\bibinfo  {journal} {Nat. Mater.}\ }\textbf {\bibinfo {volume} {25}},\ \bibinfo {pages} {884} (\bibinfo {year} {2026})}\BibitemShut {NoStop}%
\bibitem [{\citenamefont {Wang}\ \emph {et~al.}(2025)\citenamefont {Wang}, \citenamefont {Wang}, \citenamefont {Fan}, \citenamefont {Zhou}, \citenamefont {Li},\ and\ \citenamefont {Wang}}]{wang2025two}%
  \BibitemOpen
  \bibfield  {author} {\bibinfo {author} {\bibfnamefont {S.}~\bibnamefont {Wang}}, \bibinfo {author} {\bibfnamefont {W.-W.}\ \bibnamefont {Wang}}, \bibinfo {author} {\bibfnamefont {J.}~\bibnamefont {Fan}}, \bibinfo {author} {\bibfnamefont {X.}~\bibnamefont {Zhou}}, \bibinfo {author} {\bibfnamefont {X.-P.}\ \bibnamefont {Li}},\ and\ \bibinfo {author} {\bibfnamefont {L.}~\bibnamefont {Wang}},\ }\bibfield  {title} {\bibinfo {title} {Two-dimensional dual-switchable ferroelectric altermagnets: altering electrons and magnons},\ }\href {https://pubs.acs.org/doi/full/10.1021/acs.nanolett.5c03483} {\bibfield  {journal} {\bibinfo  {journal} {Nano Lett.}\ }\textbf {\bibinfo {volume} {25}},\ \bibinfo {pages} {14618} (\bibinfo {year} {2025})}\BibitemShut {NoStop}%
\bibitem [{\citenamefont {Dong}\ \emph {et~al.}(2026)\citenamefont {Dong}, \citenamefont {Liu}, \citenamefont {Dai}, \citenamefont {Guo}, \citenamefont {Xiang},\ and\ \citenamefont {Gong}}]{Dong2025}%
  \BibitemOpen
  \bibfield  {author} {\bibinfo {author} {\bibfnamefont {M.~Q.}\ \bibnamefont {Dong}}, \bibinfo {author} {\bibfnamefont {B.}~\bibnamefont {Liu}}, \bibinfo {author} {\bibfnamefont {Z.~H.}\ \bibnamefont {Dai}}, \bibinfo {author} {\bibfnamefont {Z.-X.}\ \bibnamefont {Guo}}, \bibinfo {author} {\bibfnamefont {H.}~\bibnamefont {Xiang}},\ and\ \bibinfo {author} {\bibfnamefont {X.-G.}\ \bibnamefont {Gong}},\ }\bibfield  {title} {\bibinfo {title} {Fractional quantum multiferroics from coupling of fractional quantum ferroelectricity and altermagnetism},\ }\href {https://doi.org/10.1103/twhq-32db} {\bibfield  {journal} {\bibinfo  {journal} {Phys. Rev. Lett.}\ }\textbf {\bibinfo {volume} {136}},\ \bibinfo {pages} {136702} (\bibinfo {year} {2026})}\BibitemShut {NoStop}%
\bibitem [{\citenamefont {Liu}\ \emph {et~al.}(2025{\natexlab{b}})\citenamefont {Liu}, \citenamefont {Zhao}, \citenamefont {Bellaiche},\ and\ \citenamefont {Ma}}]{fg6w-c6fd}%
  \BibitemOpen
  \bibfield  {author} {\bibinfo {author} {\bibfnamefont {X.}~\bibnamefont {Liu}}, \bibinfo {author} {\bibfnamefont {H.~J.}\ \bibnamefont {Zhao}}, \bibinfo {author} {\bibfnamefont {L.}~\bibnamefont {Bellaiche}},\ and\ \bibinfo {author} {\bibfnamefont {Y.}~\bibnamefont {Ma}},\ }\bibfield  {title} {\bibinfo {title} {Ferroelectrically switchable anomalous $\mbox{Hall}$ conductivity and nonlinear {Drude} conductivity in multiferroics},\ }\href {https://doi.org/10.1103/fg6w-c6fd} {\bibfield  {journal} {\bibinfo  {journal} {Phys. Rev. Lett.}\ }\textbf {\bibinfo {volume} {135}},\ \bibinfo {pages} {216801} (\bibinfo {year} {2025}{\natexlab{b}})}\BibitemShut {NoStop}%
\bibitem [{\citenamefont {Zhao}\ \emph {et~al.}(2024)\citenamefont {Zhao}, \citenamefont {Tao}, \citenamefont {Fu}, \citenamefont {Bellaiche},\ and\ \citenamefont {Ma}}]{PhysRevLett.133.096802}%
  \BibitemOpen
  \bibfield  {author} {\bibinfo {author} {\bibfnamefont {H.~J.}\ \bibnamefont {Zhao}}, \bibinfo {author} {\bibfnamefont {L.}~\bibnamefont {Tao}}, \bibinfo {author} {\bibfnamefont {Y.}~\bibnamefont {Fu}}, \bibinfo {author} {\bibfnamefont {L.}~\bibnamefont {Bellaiche}},\ and\ \bibinfo {author} {\bibfnamefont {Y.}~\bibnamefont {Ma}},\ }\bibfield  {title} {\bibinfo {title} {General theory for longitudinal nonreciprocal charge transport},\ }\href {https://doi.org/10.1103/PhysRevLett.133.096802} {\bibfield  {journal} {\bibinfo  {journal} {Phys. Rev. Lett.}\ }\textbf {\bibinfo {volume} {133}},\ \bibinfo {pages} {096802} (\bibinfo {year} {2024})}\BibitemShut {NoStop}%
\bibitem [{\citenamefont {Urru}\ \emph {et~al.}(2025)\citenamefont {Urru}, \citenamefont {Seleznev}, \citenamefont {Teng}, \citenamefont {Park}, \citenamefont {Reyes-Lillo},\ and\ \citenamefont {Rabe}}]{v3fg-6smc}%
  \BibitemOpen
  \bibfield  {author} {\bibinfo {author} {\bibfnamefont {A.}~\bibnamefont {Urru}}, \bibinfo {author} {\bibfnamefont {D.}~\bibnamefont {Seleznev}}, \bibinfo {author} {\bibfnamefont {Y.}~\bibnamefont {Teng}}, \bibinfo {author} {\bibfnamefont {S.~Y.}\ \bibnamefont {Park}}, \bibinfo {author} {\bibfnamefont {S.~E.}\ \bibnamefont {Reyes-Lillo}},\ and\ \bibinfo {author} {\bibfnamefont {K.~M.}\ \bibnamefont {Rabe}},\ }\bibfield  {title} {\bibinfo {title} {{$G$}-type antiferromagnetic {$\mathrm{BiFeO}_{3}$} is a multiferroic $g$-wave altermagnet},\ }\href {https://doi.org/10.1103/v3fg-6smc} {\bibfield  {journal} {\bibinfo  {journal} {Phys. Rev. B}\ }\textbf {\bibinfo {volume} {112}},\ \bibinfo {pages} {104411} (\bibinfo {year} {2025})}\BibitemShut {NoStop}%
\bibitem [{\citenamefont {Eto}\ and\ \citenamefont {Knolle}(2026)}]{eto2026odd}%
  \BibitemOpen
  \bibfield  {author} {\bibinfo {author} {\bibfnamefont {R.}~\bibnamefont {Eto}}\ and\ \bibinfo {author} {\bibfnamefont {J.}~\bibnamefont {Knolle}},\ }\bibfield  {title} {\bibinfo {title} {Odd-parity magnons in the $\mbox{Haldane-Hubbard}$ model from topological exciton condensation},\ }\href {https://arxiv.org/abs/2606.05837} {\bibfield  {journal} {\bibinfo  {journal} {arXiv:2606.05837}\ } (\bibinfo {year} {2026})}\BibitemShut {NoStop}%
\bibitem [{\citenamefont {Zhang}\ \emph {et~al.}(2026)\citenamefont {Zhang}, \citenamefont {Xie}, \citenamefont {Yu}, \citenamefont {Liu},\ and\ \citenamefont {Liu}}]{zhang2026odd}%
  \BibitemOpen
  \bibfield  {author} {\bibinfo {author} {\bibfnamefont {P.}~\bibnamefont {Zhang}}, \bibinfo {author} {\bibfnamefont {S.-B.}\ \bibnamefont {Xie}}, \bibinfo {author} {\bibfnamefont {J.}~\bibnamefont {Yu}}, \bibinfo {author} {\bibfnamefont {Y.}~\bibnamefont {Liu}},\ and\ \bibinfo {author} {\bibfnamefont {C.-C.}\ \bibnamefont {Liu}},\ }\bibfield  {title} {\bibinfo {title} {Odd-parity magnons},\ }\href {https://arxiv.org/abs/2605.31411} {\bibfield  {journal} {\bibinfo  {journal} {arXiv:2605.31411}\ } (\bibinfo {year} {2026})}\BibitemShut {NoStop}%
\end{thebibliography}%

\end{document}